\documentclass{emulateapj}
\newcommand{\Nsystab}{10}
\newcommand{\Nsysanal}{2}

\newcommand{\Nsysdisc}{2}
\newcommand{\Npldisc}{4}

\newcommand{\KepNumXXX}{23}
\newcommand{\KepNumYYY}{24}

\newcommand{\etal}{{et al.\ }}
\newcommand{\Kepler}{{\sl Kepler}\ }
\newcommand{\be}{\begin{equation}}
\newcommand{\ee}{\end{equation}}
\newcommand{\bea}{\begin{eqnarray}}
\newcommand{\eea}{\end{eqnarray}}

\slugcomment{ApJ, in press}
\shorttitle{Confirmation of Two Multiple Planet Systems}
\shortauthors{Ford et al.}

\begin{document}
\title{
Transit Timing Observations from {\em Kepler}:
II. Confirmation of Two Multiplanet Systems via a
Non-parametric Correlation Analysis
}
\author{
Eric B. Ford\altaffilmark{1}, 
Daniel C. Fabrycky\altaffilmark{2,3}, 
Jason H. Steffen\altaffilmark{4},   
%
Joshua A. Carter\altaffilmark{5,3},   
Francois Fressin\altaffilmark{5}, 
Matthew J. Holman\altaffilmark{5},  
Jack J. Lissauer\altaffilmark{6},   
Althea V. Moorhead\altaffilmark{1}, 
Robert C. Morehead\altaffilmark{1}, 
Darin Ragozzine\altaffilmark{5},    
Jason F. Rowe\altaffilmark{6,7},    
William F. Welsh\altaffilmark{8},  
Christopher Allen\altaffilmark{9}, 
Natalie M. Batalha\altaffilmark{10}, 
William J. Borucki\altaffilmark{6}, 
Stephen T. Bryson\altaffilmark{6},  
Lars A. Buchhave\altaffilmark{11,12},  
Christopher J. Burke\altaffilmark{6,7}, 
Douglas A. Caldwell\altaffilmark{6,7},  
David Charbonneau\altaffilmark{5}, 
Bruce D. Clarke\altaffilmark{6,7},    
William D. Cochran\altaffilmark{13}, 
Jean-Michel D\'{e}sert\altaffilmark{5}, 
Michael Endl\altaffilmark{13},  
Mark E. Everett\altaffilmark{14}, 
Debra A. Fischer\altaffilmark{15}, 
Thomas N. Gautier III\altaffilmark{16}, 
Ron L. Gilliland\altaffilmark{17},  
Jon M. Jenkins\altaffilmark{6,7},  
Michael R. Haas\altaffilmark{6}, 
Elliott Horch\altaffilmark{18}, 
Steve B. Howell\altaffilmark{6}, 
Khadeejah A. Ibrahim\altaffilmark{9}, 
Howard Isaacson\altaffilmark{19}, 
David G. Koch\altaffilmark{6}, 
David W. Latham\altaffilmark{5}, 
Jie Li\altaffilmark{6,7},  
Philip Lucas\altaffilmark{20}, 
Phillip J. MacQueen\altaffilmark{13}  
Geoffrey W. Marcy\altaffilmark{19}, 
Sean McCauliff\altaffilmark{9},  
Fergal R. Mullally\altaffilmark{6,7},  
Samuel N. Quinn\altaffilmark{5},  
Elisa Quintana\altaffilmark{6,7}, 
Avi Shporer\altaffilmark{21,22}, 
Martin Still\altaffilmark{6,23}, 
Peter Tenenbaum\altaffilmark{6,7},  
Susan E. Thompson\altaffilmark{6,7},   
Guillermo Torres\altaffilmark{5},   
Joseph D. Twicken\altaffilmark{6,7}, 
Bill Wohler\altaffilmark{9}, 
and the \Kepler Science Team. 
}
\altaffiltext{1}{Astronomy Department, University of Florida, 211 Bryant Space Sciences Center, Gainesville, FL 32611, USA}
\altaffiltext{2}{UCO/Lick Observatory, University of California, Santa Cruz, CA 95064, USA}
\altaffiltext{3}{Hubble Fellow}
\altaffiltext{4}{Fermilab Center for Particle Astrophysics, P.O. Box 500, MS 127, Batavia, IL 60510}
\altaffiltext{5}{Harvard-Smithsonian Center for Astrophysics, 60 Garden Street, Cambridge, MA 02138, USA}
\altaffiltext{6}{NASA Ames Research Center, Moffett Field, CA, 94035, USA}
\altaffiltext{7}{SETI Institute, Mountain View, CA, 94043, USA}
\altaffiltext{8}{Astronomy Department, San Diego State University, San Diego, CA 92182-1221 , USA}
\altaffiltext{9}{Orbital Sciences Corporation/NASA Ames Research Center, Moffett Field, CA 94035, USA}
\altaffiltext{10}{Department of Physics and Astronomy, San Jose State University, San Jose, CA 95192, USA}
\altaffiltext{11}{Niels Bohr Institute, Copenhagen University, DK-2100 Copenhagen, Denmark}
\altaffiltext{12}{Centre for Star and Planet Formation, Natural History Museum of Denmark, University of Copenhagen, DK-1350 Copenhagen, Denmark}
\altaffiltext{13}{McDonald Observatory, The University of Texas, Austin TX 78730, USA}
\altaffiltext{14}{National Optical Astronomy Observatory, Tucson, AZ 85719, USA}
\altaffiltext{15}{Department of Astronomy, Yale University, New Haven, CT 06511, USA}
\altaffiltext{16}{Jet Propulsion Laboratory/California Institute of Technology, Pasadena, CA 91109, USA}
\altaffiltext{17}{Space Telescope Science Institute, Baltimore, MD 21218, USA}
\altaffiltext{18}{Department of Physics, Southern Connecticut State University, New Haven, CT 06515, USA}
\altaffiltext{19}{ Astronomy Department, University of California, Berkeley, Berkeley, CA 94720}
\altaffiltext{20}{Centre for Astrophysics Research, University of Hertfordshire, College Lane, Hatfield, AL10 9AB, England}
\altaffiltext{21}{Las Cumbres Observatory Global Telescope, Goleta, CA 93117, USA}
\altaffiltext{22}{Department of Physics, Broida Hall, University of California, Santa Barbara, CA 93106, USA}
\altaffiltext{23}{Bay Area Environmental Research Institute/NASA Ames Research Center, Moffett Field, CA 94035, USA}
\email{eford@astro.ufl.edu}

\begin{abstract}
We present a new method for confirming transiting planets based on the combination of transit timingn variations (TTVs) and dynamical stability.  
Correlated TTVs provide evidence that the pair of bodies are in the same physical system.  
Orbital stability provides upper limits for the masses of the transiting companions that are in the planetary regime.  
This paper describes a non-parametric technique for quantifying the statistical significance of TTVs based on the correlation of two TTV data sets.  
We apply this method to an analysis of the transit timing variations of two stars with multiple transiting planet candidates identified by Kepler.  
We confirm four transiting planets in two multiple planet systems based on their TTVs and the constraints imposed by dynamical stability.  
An additional three candidates in these same systems are not confirmed as planets, but are likely to be validated as real planets once further observations and analyses are possible.  If all were confirmed, these systems would be near 4:6:9 and 2:4:6:9 period commensurabilities.  
Our results demonstrate that TTVs provide a powerful tool for confirming transiting planets, including low-mass planets and planets around faint stars for which Doppler follow-up is not practical with existing facilities.  Continued Kepler observations will dramatically improve the constraints on the planet masses and orbits and provide sensitivity for detecting additional non-transiting planets.  
If Kepler observations were extended to eight years, then a similar analysis could likely confirm systems with multiple closely spaced, small transiting planets in or near the habitable zone of solar-type stars.  
\end{abstract}


\section{Introduction} 
\label{secIntro}
%
%
NASA's \Kepler spacecraft was launched on March 6, 2009, with the goal of characterizing the occurrence of small exoplanets around solar-type stars.  The nominal \Kepler mission was designed to search for small planets transiting their host stars by observing targets spread over 100 square degrees for three and a half years (Borucki \etal 2010; Koch \etal 2010).  
%
%
After an initial period (``quarter'' 0; Q0) of collecting engineering data that included  observations of a subset of the planet search stars, \Kepler began collecting science data for over 150,000 stars on May 13, 2009.  On February 1, 2011, the \Kepler team released light curves observed during the first $\sim4.5$ months of observations (Q0, Q1 and Q2; May 2, 2009-September 16, 2009) for all planet search targets via the Multi-Mission Archive at the Space Telescope Science Institute (MAST; \url{http://archive.stsci.edu/kepler/}).  
In this paper, we analyze observations of \Nsysanal\ stars taken through the end of Q6 (September 22, 2010) which will be made publicly available from MAST at the time of publication.  

%
%
Borucki \etal (2011; hereafter B11) reported the results of an initial search for transiting planets based on an early and incomplete version of the \Kepler pipeline.  B11 identified 1235 Kepler Objects of Interest (KOIs) that were active planet candidates as of 1 February 2011 and had transit-like events observed in Q0-2.  For each of these, B11 Table 2 lists the putative orbital period, transit epoch, transit duration, planet size, and other properties.  B11 Table 1 lists properties of the host star, largely from pre-launch, ground-based photometry obtained in order to construct the \Kepler Input Catalog (KIC; Brown \etal 2011).  
Additional candidates will likely be identified due to improvements in the \Kepler pipeline and the availability of additional data.

Many KOIs were quickly recognized as likely astrophysical false positives (e.g., blends with background eclipsing binaries; EBs) and were reported in B11 Table 4.  
For the remaining planet candidates, the \Kepler team reported a ``vetting flag'' in B11 to indicate which KOIs are the strongest and weakest planet candidates.  B11 estimates the reliability to be: $\ge$98\% for confirmed planets (vetting flag=1), 
$\sim80\%$ for strong candidates (vetting flag=2), and $\sim60\%$ for less strong candidates (vetting flag=3) or candidates that have yet to be vetted (vetting flag=4).  Independent calculations suggest that the reliability could be even greater (Morton \& Johnson 2011).  

%
%
Further follow-up observations and analysis are required to determine the completeness and false alarm rates as a function of planet and host star properties.  Nevertheless, several papers have begun to analyze the properties of the \Kepler planet candidate sample.  Both B11 and Howard \etal (2011) assumed that most \Kepler planet candidates are real and analyzed the frequency of planets as a function of planet size, orbital period and host star type.  Youdin (2011) performed a complementary analysis of the joint planet size-orbital period distribution.  Moorhead \etal (2011) analyzed the transit duration distribution of \Kepler planet candidates and the implications for their eccentricity distribution.  
%
%
Of particular interest are 115 stars, each with multiple transiting planet candidates (MTPCs).  
Latham \etal (2011) compared the planet size and period distributions of these planet candidates with those orbiting stars with only one transiting planet candidate.  Lissauer \etal (2011b) analyzed the architectures of such planetary systems.  Ford \etal (2011) 
searched for evidence of transit timing variations (TTVs) based on the transit times measured during Q0-2.  They found 65 TTV candidates and identified a dozen MTPC systems which are likely to be confirmed (or rejected)by TTVs measured over the full \Kepler mission lifetime.  
Both Latham \etal (2011) and Lissauer \etal (2011b) point out that the number of false positives in MTPC systems is expected to be much lower than the number of false positives for single transiting planet candidates.  Combined with the already low rate of false positives for \Kepler planet candidates (B11; Morton \& Johnson 2011), we expect very few false positives among candidates in MTPC systems.  Still, it is important to test and confirm individual systems, in order to establish the frequency of planetary systems with small planets and identify any unexpected sources of false positives.
 
Already, seven \Kepler planets have been confirmed by TTVs (Kepler-9b\&c, Holman \etal 2010; Kepler-11b-f, Lissauer \etal 2011a) and three additional \Kepler planets in MTPC systems have been validated (Kepler-9d, Torres \etal 2011; Kepler-10c, Fressin \etal 2011; Kepler-11g, Lissauer \etal 2011).  Previously, discovery papers presented a detailed analysis of all available data for each confirmed \Kepler planet, often including a wide array of follow-up observations.  Given {\em Kepler}'s astounding haul of planet candidates, such detailed analyses will not be practical for all planets.  

In this paper, we present a new method to confirm planets based on combining observations of TTVs with the constraint of dynamical stability.  We perform TTV analyses of \Nsysdisc~MTPC systems that provide strong evidence that at least two of the planet candidates around each star are bound to the same host star (as opposed to a blend of two stars each with one planet or two eclipsing binaries diluted by the target star).  Next, we test for dynamical stability of the nominal multiple planet system model and consider the effect of varying the mass of the planet candidates.  We place upper limits on the masses of planets which show significant TTVs.  
Finally, we perform a basic analysis of \Kepler observations through Q6 and the available follow-up observations to check for any warning signs of possible false positives.  
%
Given the 
evidence for TTVs and the mass limits from dynamical stability, we confirm \Npldisc~planets in \Nsysdisc~MTPC systems.  

%
%
This paper is organized as follows.  First, we provide an overview of \Kepler observations for the KOIs considered in \S\ref{secObs}.   
Second, we describe a new method for calculating the significance of TTVs in MTPC systems in \S\ref{secStatMethods}-\ref{secSignificance}.  
Readers primarily interested in the properties of the planetary systems do not necessarily need to read the details of the statistical methods presented in \S\ref{secGP}-\ref{secCor}.  
We describe our use of n-body simulations to obtain upper limits on planet masses in \S\ref{secNbodyMethods}.
In \S\ref{secResults}, we describe the results of our statistical and dynamical analyses.
We present available follow-up observations and additional analysis of \Kepler data in \S\ref{secFop} and discuss each system individually in \S\ref{secSysProp}. 
Finally, we discuss the implications of our results and prospects of the method for the future in \S\ref{secDiscuss}.

\section{\Kepler Photometry \& Transit Times}
\label{secObs}
%
\label{secTTs}
We measured transit times based on the ``corrected'' (or PDC) long cadence (LC), optimal aperture photometry performed by the \Kepler Science Operations Center (SOC) pipeline version 6.2 (Jenkins \etal 2010).  
The details of the properties of PDC data and how we deal with the most common complications are described in Ford \etal (2011).  For each system, we begin using the same procedure to measure transit times in bulk.  
In short, for each planet candidate we fold light curves at an initial estimate of the orbital period.  For each planet candidate, we fit a transit model to the folded light curve, excluding observations during the transits of other planet candidates.  We use the best-fit limb-darkened transit model as a fixed template when fitting for each transit time.  For example, for a three planet system, we first remove transits of KOIs .02 and .03 from the original light curve to measure TTs for KOI .01.  Then, we remove transits of KOIs .01 and .03 from the light curve to measure TTs for KOI .02.  Finally, we remove transits of KOIs .02 and .03 to measure TTs for KOI .01.  In some cases, we iterate the procedure.  This consists of aligning the transits based on the initial set of measured TTs to generate a new transit template and remeasuring transit times using the new template.  The choice of which TTs to measure and which to exclude due to data gaps or anomalies may also change upon iteration.  See Ford \etal (2011) for details about the algorithm for measuring transit times.  We estimate the uncertainty in each transit time from the covariance matrix.  

For some planet candidates with shallow transits, each transit is too shallow for individual transits to be clearly detected, and a new algorithm by J. Carter was employed.  This model attempts to fit only data within four transit durations of the suspected mid-transit, and allows for a linear baseline, to correct for astrophysical and instrumental drifts on timescales much longer than the transit duration.  Global transit parameters are solved for to create a template light curve, which is then scanned over each individual transit.  In that scanning process, $\chi^2$ is densely sampled as a function of the proposed transit mid-time.  Because of the low SNR, the shape of $\chi^2$ as a function of the midtime can be very skewed and spiky, rather than shaped as a parabola, in the vicinity of the local minimum.  Therefore, instead of the local curvature for an error bar, the algorithm fits a parabola to the sampled $\chi^2$ function, out to $\Delta \chi^2=7$ away from the minimum.  This parabola thus has a width which is more stable to noise properties than the local curvature is, and we adopt its width as the error bar in each point.  We employed this method for KOIs 168.03, 1102.01 and 1102.02 that are described in detail in \S\ref{secSysProp}.

We report best-fit linear ephemerides based on transits observed during Q0-6 in Table \ref{tabPlanets}, along with the number of transit times observed (nTT), the median timing uncertainty ($\sigma_{TT}$) and the median absolute deviation from the linear ephemeris (MAD).  
Transit times measured are reported in Table \ref{tabTTs}.   By definition, positive TTs occur later than predicted by the linear ephemeris.

\section{Methods}
\label{secMethods}

\subsection{Statistical Analysis}
\label{secStatMethods}
Visual inspection of TT data sets revealed KOIs with multiple transiting planet candidates including two planet candidates which appear to have TTVs that are anticorrelated with each other.  In this and two companion papers (Fabrycky \etal 2011; Steffen \etal 2011), we develop complementary methods to establish the statistical significance of apparently correlated TTVs measured in systems of multiple transiting planet candidates.  Physically, transit timing variations (relative to a linear ephemeris) can only be  measured at the times of transit.  Since transit times can only be measured at discrete times and transits of two planets are rarely coincident with each other, one can not calculate the standard correlation coefficient, since that would require data sampled either continuously or at common times.  In this section we develop tools to quantify the extent of the correlations between TTV curves and the statistical significance of the TTVs in MTPC systems.  

Even if we restrict our attention to systems with only two planet candidates, there is an astounding variety of potential TTV signatures (e.g., Veras \etal 2011).  For some systems (e.g., in a 1:2 mean motion resonance (MMR) with small libration amplitude), the TTV signature may be relatively simple (e.g., nearly sinusoidal) for the timespans and timing precision of interest.  In these cases, one might be well served by fitting a parametric model to the observations (e.g., polynomial, sinusoid).  However, for other systems (e.g., slightly offset from resonance, modest eccentricity, more than two planets), the TTV signature can be quite complex.  Often the shape and amplitude of the TTVs changes from year to year (or even longer timescales; e.g., Fig 2. of Ford \& Holman 2007).  Given the diversity and potential complexity of TTV signatures, it is necessary to consider a broad range of functional forms and a large number of model parameters.  Normally, this would raise concerns about exploring the high-dimensional parameter space and potentially over-fitting data.  

On a simplistic level, one could address this problem by smoothing the TTV observations to obtain a continuous ``TTV curve'' for each planet candidate and calculate the standard correlation coefficient between the two smoothed TTV curves.  Of course, the results will depend on the choice of smoothing algorithm.  We overcome the challenges of working with a discrete dataset with potentially complex structure by the application of Gaussian Processes (GPs).  While one could think of the GPs as a fancy smoothing algorithm, there are several attractive features of GPs that make them well suited for this application.  In particular, GPs are infinite dimensional objects, providing them with enormous flexibility for modeling the data.  At the same time, their mathematical properties make it practical to marginalize over the infinite dimensional parameter space to calculate, not just the predictions of a GP for the TTV curve, but also the full posterior predictive probability distribution for the TTV curve at any finite number of times (see \S\ref{secGPmethod}). Thus, we are able to naturally account for the uncertainties in the model TTV curve in a fully Bayesian manner.  

Once we obtain a continuous function for the ``TTV curve'' via our GP model, we calculate a Pearson correlation coefficient between the GP models for each pair of neighboring planets.  To establish the statistical significance of the apparently correlated TTVs, we calculate the distribution of correlation coefficients calculated similarly, but applied to synthetic data sets.  Each synthetic data set is a random permutation of the actual data.  If the correlation coefficient calculated from the actual data is more extreme than the 99.9th percentile of the correlation coefficients calculated from the simulated data, then we conclude that TTVs are sufficiently statistically significant to consider the planet pair confirmed.

\subsection{Overview of Gaussian Processes}
\label{secGP}

Readers who are not interested in the details of the details of the statistical
methods used to establish the significance of TTVs in MTPC systems may choose to skip \S\ref{secGP}-\ref{secCor}.  
GPs have been extensively studied and applied in the statistics and machine learning communities (Rasmussen \& Williams 2006).  GPs have also been applied in a variety of astronomical contexts (e.g., Rybicki \etal 1992; Gibson \etal 2011). A GP defines a distribution over functions.  ``Process'' refers to a method for generating a time series of data.  ``Gaussian'' refers to the defining property of a GP that the joint distribution of any finite number of measurements (at discrete times) has a multi-variate Gaussian distribution, even when conditioned on any combination of observations.  While this assumption may seem to be restrictive, GPs are extremely versatile.  The assumption of Gaussianity is essential for making it practical to perform computations on the infinite dimensional function space.  In particular, for a GP that describes functions of time only ($f(t)$), a particular GP ($\mathcal{GP}\left[m(t),k(t,t')\right]$)is specified by a mean value, $m(t)$, and a covariance function, $k(t,t')$.  Once we adopt a form for the covariance function (i.e., for particular values for hyperparameters; see \S\ref{secGPmethod}), it becomes practical to perform Bayesian inference and calculate the posterior predictive probability distribution for the values of $f(t)$ at any number of times.  
%
%

For our application, we can intuitively think of TTVs as measurements of the value of a function ($f(t)$) that is related to how much a planet is ahead (or behind) schedule in its orbit.
\footnote{
%
An alternative interpretation would define a continuum of hypothetical observers distributed along the orbital plane.
}
Of course, \Kepler can only make measurements of this function at times of transit (or occultation) as viewed by {\sl Kepler}.  Using a Bayesian framework, we can calculate the posterior probability distribution for $f(\mathbf{t}^*)$ at hypothetical observation times ($t^*_i$), conditioned on the actual measurements of $f(\mathbf{t})$ (i.e., TTVs) and hyperparameters ($\mathbf{\theta}$) that specify the form of the covariance function.  
We perform this procedure for each planet candidate in a system.  Then, we can calculate the correlation coefficient of the GPs for a pair of transiting planet candidates.  We focus our attention on neighboring pairs of transiting planet candidates, meaning we do not search for significant correlations between planets if there is an additional transiting planet candidate with an intermediate period.  In order to establish the statistical significance of the correlation coefficient, we perform the same procedure on synthetic data sets, generated by permuting the order of the TTVs (along with their measurement uncertainties).  We compare the correlation coefficient for the actual TTVs to the distribution of correlation coefficients for the synthetic data sets to determine a false alarm probability for the existence of TTVs.

\subsection{Gaussian Process Model}
\label{secGPmethod}

We model the TTVs of each KOI as an independent Gaussian Process (GP).  
Following Rasmussen \& Williams (2006), the prior probability distribution for the values at times ($\mathbf{t^*}$) of a GP with zero mean is 
\be
f_* \equiv f(t^*) \sim \mathcal{N}\left(\mathbf{0},K\left(t^*,t^*\right)\right),
\ee
where $K(t^*,t^*)$ is the correlation matrix.  
We assume that each observed transit time, $y_i$, is normally distributed about the true transit time ($f(t_i)$) and that independent measurement errors have a variance of $\sigma_{obs,i}^2$.  Then the joint distribution for the actual observations and a set of hypothetical measurements is
\be
\label{eqnGp1}
\left[\begin{array}{c} \mathbf{y} \\ \mathbf{f^*} \end{array} \right] 
\sim \mathcal{N}\left(\mathbf{0}, 
\left[\begin{array}{cc} 
K\left(\mathbf{t},\mathbf{t}\right)+\mathbf{\sigma_{obs}}^2 &  
K\left(\mathbf{t},\mathbf{t^*}\right) \\
K\left(\mathbf{t^*},\mathbf{t}\right) &
K\left(\mathbf{t^*},\mathbf{t^*}\right)
\end{array} \right] 
\right),
\ee
where $\mathbf{\sigma}^2_{obs}$ is a diagonal matrix with entries of $\sigma_{obs,i}^2$.
We can calculate the posterior predictive distribution by conditioning the joint prior distribution on the actual observations, $\mathbf{y}$.  The standard results for the mean ($\bar{\mathbf{f}}^*$) and covariance ($\mathrm{cov}(\mathbf{f^*}))$ of the posterior predictive distribution are
\bea
\label{eqnGpPred}
\bar{\mathbf{f}}^* & = & K(\mathbf{t^*},\mathbf{t})\left[K(\mathbf{t},\mathbf{t})+\mathbf{\sigma}^2_{obs}\right]^{-1} \mathbf{y} \\
\mathrm{cov}(\mathbf{f^*}) & = & K(\mathbf{t^*},\mathbf{t^*}) - K(\mathbf{t^*},\mathbf{t}) \left[K(\mathbf{t},\mathbf{t})+\mathbf{\sigma}^2_{obs}\right]^{-1} K(\mathbf{t},\mathbf{t^*}).
\eea
%

For our calculations, we adopt a Gaussian covariance matrix
\be
K(t_i,t_j; \sigma_r, \tau) = \sigma_r^2 \exp\left(-\frac{\left(t_i-t_j\right)^2}{2\tau^2}\right),
\ee
where $\sigma_r$ and $\tau$ are hyperparameters, describing the amplitude and timescale of correlations among data points.  This choice of a kernel function ensures that the Gaussian process is a smooth function (i.e., continuously differentiable) and allows there to be a single characteristic timescale for TTVs.  It can be shown that covariance matrices of this form correspond to Bayesian linear regression model with an infinite number of Gaussian basis functions (Rasmussen \& Williams 2006).  

In modeling the TTVs as a GP, we are making use of a posterior predictive distribution for all functions that are consistent with the observations and the covariance matrix specified by a set of hyperparameters $\mathbf{\theta}$.  The ``trick'' of GPs is that we can marginalize over all functions analytically using matrix algebra.  
Fortunately, the log marginal likelihood conditioned on the hyperparameters can also be readily calculated via matrix algebra
\bea
\log p(\mathbf{y}|\mathbf{t}, \mathbf{\theta}) & = & -\frac{1}{2}\mathbf{y}^T \left(K_\theta+\mathbf{\sigma}^2_{obs}\right)^{-1}\mathbf{y} \\
& & -\frac{1}{2}\log\left|K_\theta+\mathbf{\sigma}^2_{obs}\right| - \frac{n}{2} \log 2\pi,
\eea
where $K_{\mathbf{\theta}}$ is the correlation matrix $K$ evaluated at the observation times, $\mathbf{t}$, using a set of hyperparameters $\mathbf{\theta} = \left\{\log \sigma_r, \log \tau \right\}$ (Rasmussen \& Williams 2006).
We set $\sigma_r^2 = \mathrm{median}(y_i^{2})$, which is equivalent to normalizing the TTV observations by their median absolute deviation.  We adopt a value of $\tau$ that maximizes the marginal likelihood.  We expect this to be a good approximation for data sets of interest, since the likelihood is typically sharply peaked for datasets with a significant structure in the TTV curve.  Technically, the marginal likelihood can be bimodal, with one mode having small $\tau$ (corresponding to functions that model the observations well) and a second mode with large $\tau$ (corresponding to functions that basically ignore the measure transit times, attributing them to measurement noise).  In principle, one could explicitly compare the marginal likelihood of each mode or even use the weighted average of the GPs corresponding to the two local maxima.  In practice, we found that this was not a problem for the data sets considered.  After verifying that the results were not sensitive to the initial guess for $\tau$, we adopted an initial guess for $\tau$ equal to the shorter orbital period of the two planets or 90 days (for long period planets).  
%
%
We use the \texttt{tnmin} minimization package provided by Craig Markwardt\footnote{\url{http://cow.physics.wisc.edu/$\sim$craigm/idl/idl.html}} that is based on the Newton method for nonlinear minimization.  We find that this algorithm is robust for the data sets considered.

We also experimented with Mat\'{e}rn class of covariance functions, which can yield GPs that are less smooth than the Gaussian covariance function.  The Mat\'{e}rn class is parameterized by both a scale $\tau$ and a new hyperparameter, $\nu$.  In the limit that $\nu\to\infty$, the Mat\'{e}rn class approaches the Gaussian covariance function.  For half-integer values of $\nu$, the Mat\'{e}rn covariance functions can be written as a polynominal times an exponetial.  In particular, we tested $\nu=5/2$, 
\bea
K(t_i,t_j; \sigma_r, \tau, \nu=\frac{5}{2}) & = & \sigma_r^2 \left(1+\frac{\sqrt{5} \Delta~t}{\tau} + \frac{5\Delta~t^2}{3\tau^2}\right)  \\
 & & \times \exp\left(-\frac{\sqrt{5}\Delta~t}{\tau}\right),
\eea
where $\Delta~t=t_i-t_j$ (Rasmussen \& Williams 2006).  While the choice of covariance function significantly affects the smoothness of the predictive distribution, the significance of the correlation between two TTV data sets did not appear to be sensitive to the choice of a Gaussian or Mat\'{e}n $\nu=5/2$ covariance function.  
Based on our experience analyzing real and simulated data sets, we found that choosing a Gaussian covariance function and the method of estimating $\tau$ described above resulted in a highly robust algorithm.  This allowed us to automate our analysis, so that we could perform the Monte Carlo simulations necessary to establish the false alarm rates.

\subsection{Correlation Coefficient}
\label{secCor}

Using the GP model for two planet candidates' TTV curves, we calculate the mean ($f_p(\mathbf{t^*})$) and variance ($\sigma^2_p(\mathbf{t^*})$) of the predictive distribution for each planet, indicated by the index $p$.  For $\mathbf{t}^*$, we adopt the observed transit times for both planets combined into a single vector.  
We calculate a weighted correlation coefficient between the two GP models based on these two samples, using weights
\be
w_{i} = \left(\sigma^2_p(t_i) + \sigma^2_{q}(t_i)\right)^{-1},
\ee
where $\sigma_p^2$ and $\sigma_q^2$ refer to the variances of the posterior predictive distribution of the Gaussian process (i.e., the diagonal elements of $\mathrm{cov}(\mathbf{f^*}$) in Eqn.\ \ref{eqnGpPred}).  After concatenating the weight vectors for the observations times of the two planets, the weighted average mean and variance of the predictive distributions are
\be
\left<f_p\right> = \left[ \sum_i w_i f_p(t_i) \right] / \sum_i w_i
\ee
and
\be
\left<\sigma_p^2\right> = \left[ \sum_i w_i \sigma^2_p(t_i) \right] / \sum_i w_i.
\ee
We calculate the covariance between the two GPs evaluated at the actual observation time according to
\be
\mathrm{cov}_{p,q} = \left[ \sum_i w_i \left( f_p(t_i) - \left<f_p\right> \right) \left( f_q(t_i) - \left<f_q\right> \right) \right] / \sum_i w_i.
\ee
Thus, the correlation coefficient between the two GPs is estimated by 
\be
C = \frac{\mathrm{cov}_{p,q}}{\sqrt{\left(\mathrm{cov}_{p,p}+\left<\sigma_p^2\right>\right)\left(\mathrm{cov}_{q,q}+\left<\sigma_q^2\right>\right)}}
\ee

\subsection{Establishing the Statistical Significance of TTVs}
\label{secSignificance}

In order to establish the statistical significance of the putative TTVs for a pair of planet candidates, we apply the same methods described in \S\ref{secGPmethod} \& \ref{secCor} to an ensemble of synthetic datasets.  Each synthetic data set includes a random permutation of the TTVs for each of the planet candidates.  The TTV and measurement uncertainty for a given transit remain paired after permutation.  
We subtract the best-fit linear ephemeris for each synthetic dataset before generating a GP model and calculating the weighted correlation coefficients ($C'_i$) for the $i$th synthetic data set.  
It is important that we subtract the best-fit linear ephemeris before generating the GP model, so that any long-term trend in the TTVs is absorbed into the best-fit orbital period.
We consider $N_{\mathrm{perm}} = 10^4$ permuted synthetic data sets.  
%
%
We estimate the false alarm probability (FAP$_{\mathrm{TTV,C}}$) based on the fraction of synthetic data sets for which $|C'|$ is greater than $\left|C\right|$ for the actual pair of TTV curves.  In cases where no synthetic data sets yield a $|C'|$ as large as $|C|$ for the actual observations, we report the FAP$_{\mathrm{TTV,C}}\le~10^{-3}$.  

In principle, the above process could result in identifying either a positive or a negative correlation coefficient.  We expect that an isolated pair of interacting planets will have a negative correlation coefficient, reflecting that energy is exchanged between the orbits.  While this is strictly true for two planet systems, one could conceive of a system with additional planets (e.g., a non-transiting planet that is perturbing both of the planets observed to transit) in which a pair of planets would have a positive correlation coefficient.  Therefore, we conservatively calculate the FAP based on the distribution of the absolute value of the correlation coefficient, rather than only the fraction of synthetic data sets with $C$ more negative than that measured for the actual observations.  

Several arguments about multiplicity suggest that systems with MTCPs have a significantly lower false alarm rate than the overall sample of \Kepler planets (Lissauer \etal 2011b).  We conservatively adopt a threshold false alarm probability for detecting TTVs of $10^{-3}$.  Using such a threshold, we expect that the probability of claiming significant TTVs for a given pair of planets due to Gaussian measurement noise is less than $10^{-3}$. 
Since B11 reported 115, 45, 8, 1 and 1 systems with two, three, four, five and six candidate transiting planets, there is a total of 238 pairs of neighboring planets which we could test (as opposed to 323 total pairs, including non-neighboring pairs).  
Thus, the expected number of statistical false alarms from the current sample of multiple planet candidate systems remains less than one, even if we were to consider every system with an FAP$<10^{-3}$ as
confirmed.  In practice, we only claim to confirm those pairs for which the FAP was robust to outliers (and passed a series of additional tests).

Our estimates of the false alarm probability assume that each transit time measurement is independent and uncorrelated with other measurements.  
%
%
While there is correlated noise in the \Kepler photometry, this does not directly translate into correlations among measured transit times, since the transits are measured at widely separated times.  One possible mechanism for generating apparently anticorrelated TTs is measurement noise due to nearly contemporaneous transits of multiple planets.  Therefore, our analysis excludes transits that are nearly coincident with the transit of another known planet candidate.  
Physically, it is extremely difficult for an alternative astrophysical process to cause the large transit timing variations observed in these systems.  Aside from the gravitational perturbations of the other transiting planet, the most plausible mechanisms would be star spots.  
However, generating TT variations on timescales longer than a year would require an unusually long-lived spot complex. Further, transit timing noise due to star spots would have an essentially random phase, except in rare cases where the planet orbital period were nearly coincident with a near integer multiple of the stellar rotation period (e.g., Desert \etal 2011a).  Such a coincidence for multiple planets in one system is even more unlikely.   Finally, the observed TTV amplitude is much greater than what can be caused by starspots (Holman \etal 2010).  As neither of the stars considered in \S\ref{secSysProp} show large rotational modulation, any starspot induced timing variations are negligible relative to TT measurement precision.  Thus, starspots are not a viable explanation of the measured TTVs for either of the stars considered in \S\ref{secSysProp}.  Indeed, if there were an autocorrelation of the TT residuals (TT observations relative to the GP model), the most likely cause would be actual timing variations due to gravitational perturbations that are not accurately described by our GP model.  Since our GP model allows for only a single timescale, an orbital configuration that results in multiple TTV timescale would naturally lead to an autocorrelation of the TT residuals.  We are optimistic that continued \Kepler observations will allow us to detect TTVs on multiple timescales, providing more precise constraints on the masses and orbits of the planets in these systems (see Fabrycky \etal 2011).


\subsection{Dynamical Analysis}
\label{secNbodyMethods}
To investigate the orbital stability of the system, we construct a nominal model based on the measured orbital periods with circular and coplanar orbits.  In the nominal model, we adopt planet masses based on the adopted planet radii, and an empirical mass-radius relationship ($M_p/M_{\oplus} = \left(R_p/R_{\oplus}\right)^{2.06}$; Lissauer et al.\ 2011ab).  We verify that the nominal model is stable for at least $10^7$ years, including all transiting planet candidates in the system.  

Second, we place upper limits on the masses of the systems with correlated TTVs 
by assuming that the real system is not dynamically unstable on a short timescale.
For placing mass limits, we start from the nominal model, but include only the pair of planets with significant TTVs.  (Including additional planet candidates would typically make the system even less likely to be stable.)  
We inflate the mass of each planet candidate by a common scale factor.  We use the same scale factor for the mass of both planets, since the relative sizes are well determined from the transit light curves (with the possible exception of grazing transits).  A misestimated stellar radius would cause both planets' sizes and hence nominal masses by a similar factor.  As a misestimated stellar radius would not significantly change the planet-planet size ratio, the nominal mass ratio for our n-body simulations is insensitive to uncertainties in the stellar radius.  
Another possibility is that both planets sizes could be significantly underestimated if the light from the target star were diluted by light from nearby stars.  Again, this could significantly affect the planet sizes, but not the planet-planet size ratio or the planet-planet mass ratio for our nominal model.  
A final possibility is that both planets with correlated TTVs are transiting a star other than the target star, which is diluted by the target star.  In this case, the planet radii would be larger than estimated by B11, but the ratio of planet radii would still be accurately estimated (assuming neither is grazing).  Importantly, in each of these cases dynamical stability would still provide upper limits for the masses that show the bodies to be planets, even if the planetary radii were significantly larger than estimated.

\section{Results \& Confirmation of \Kepler Planets}
\label{secResults}

\subsection{TTV Analysis \& Evidence for Multi-body Systems}
\label{secResultsTtv}

\label{secSysConfirm}
\begin{figure*}
\epsscale{1.0}
\plottwo{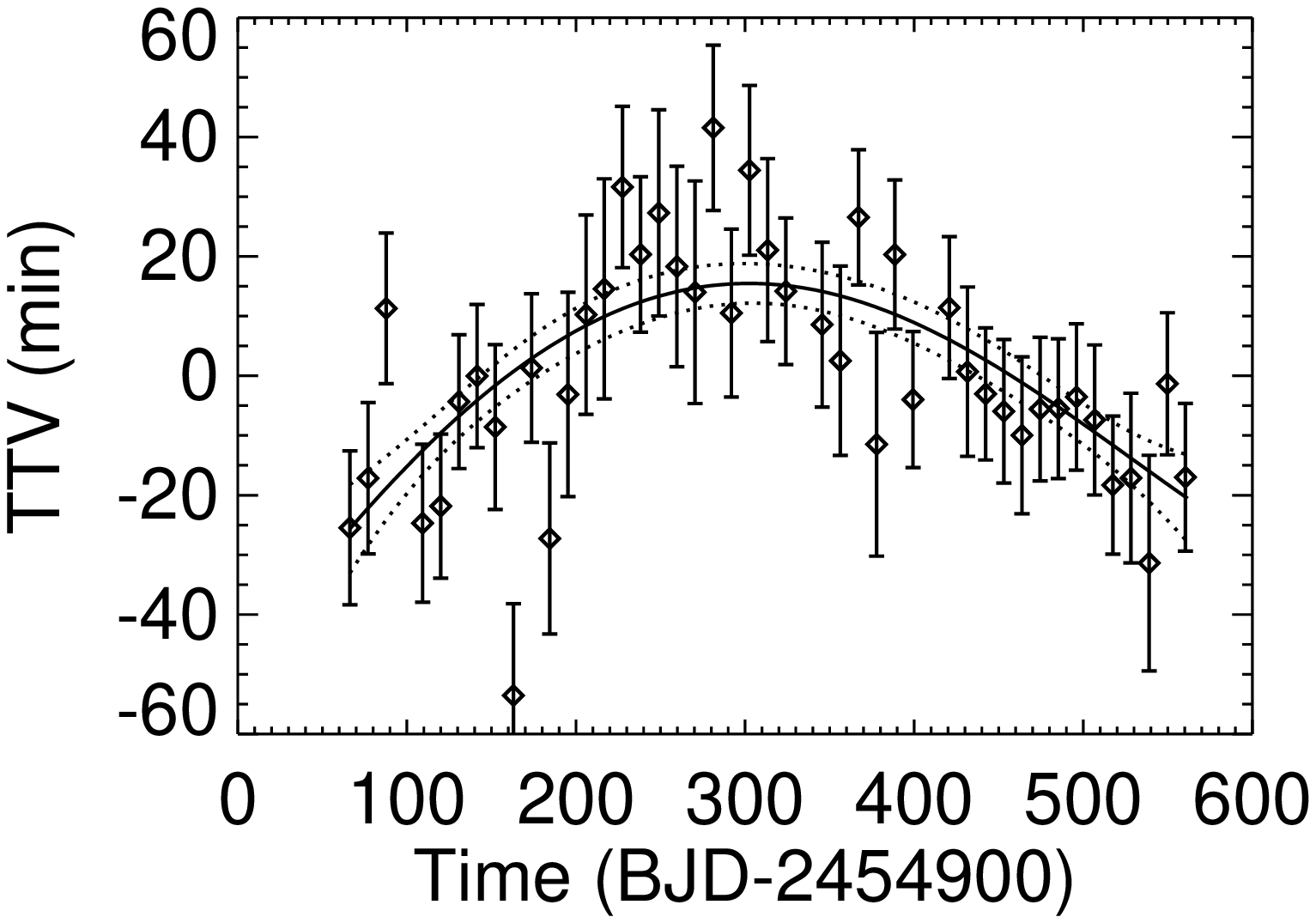}{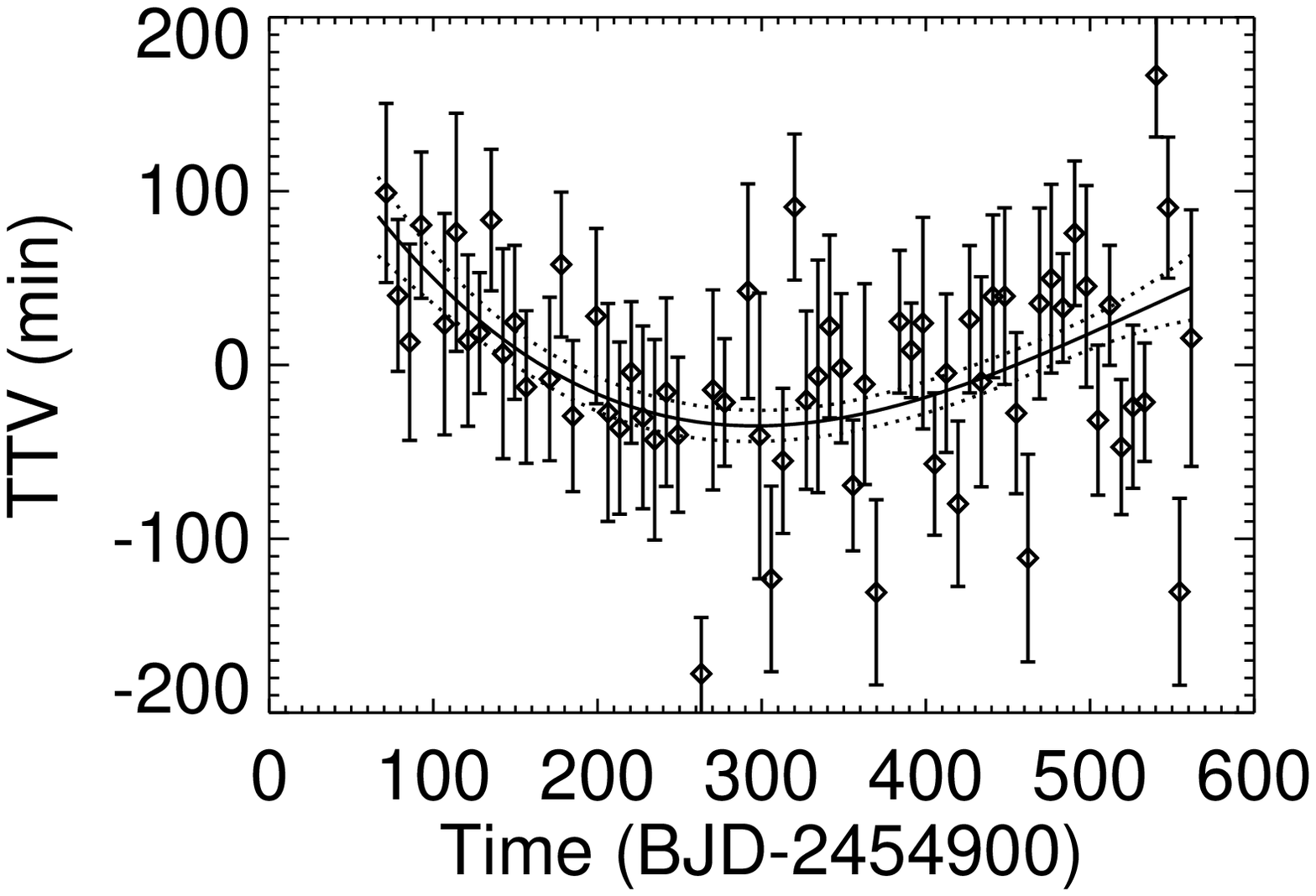}
\plottwo{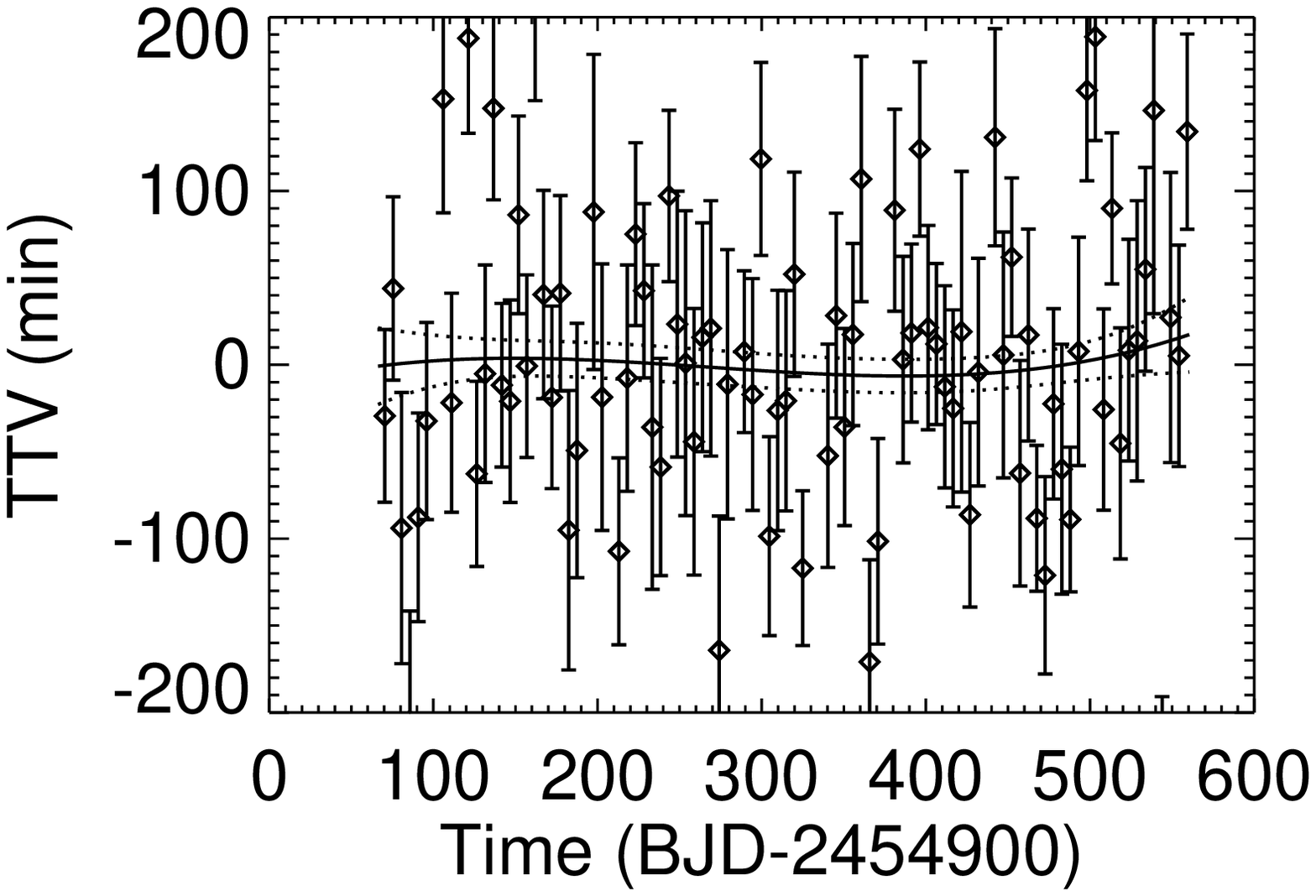}{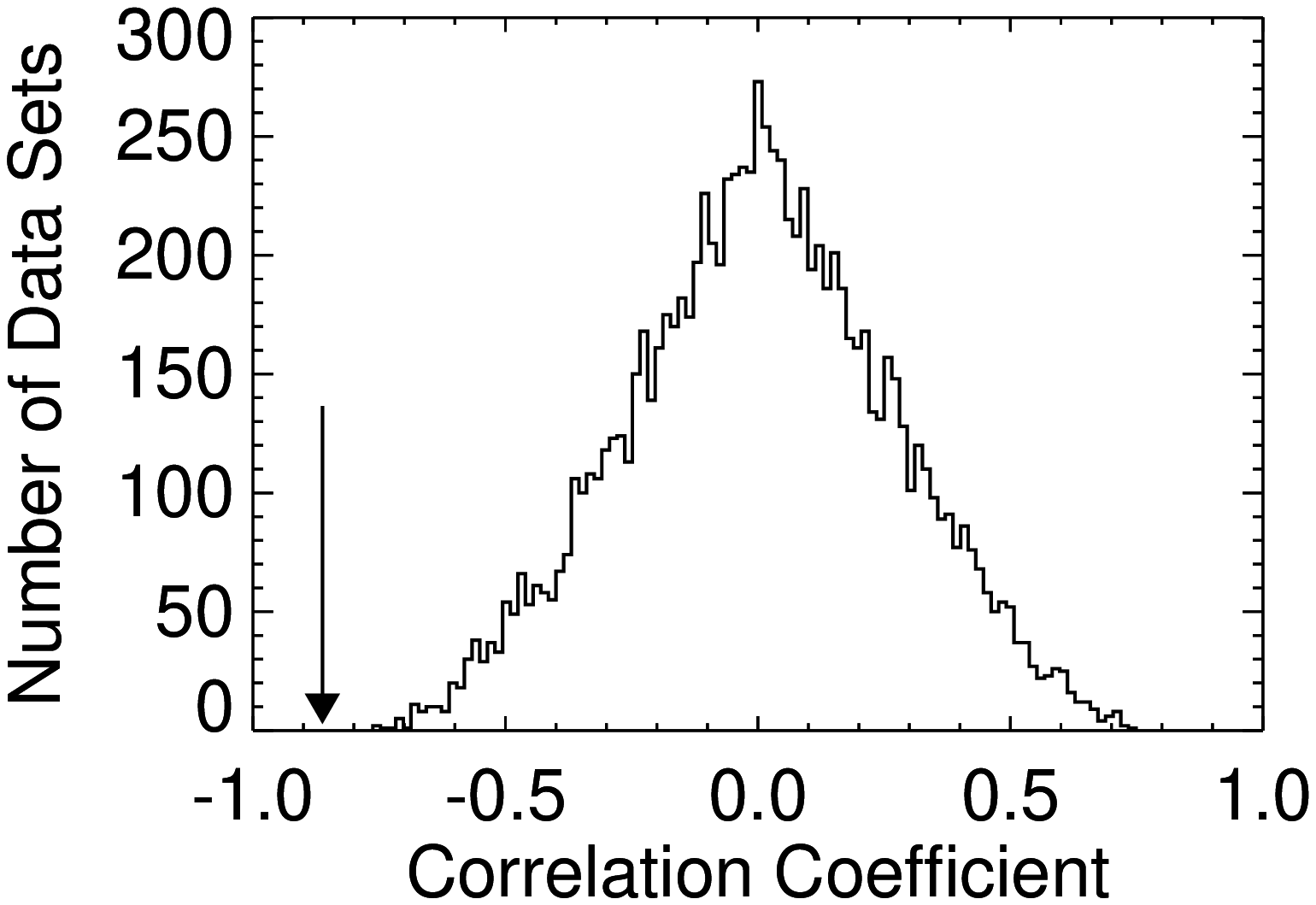}
\caption{Gaussian Process models for transit times of Kepler-23's planet candidates.  Points with error bars show deviations of measured transit times from a linear ephemeris for Kepler-23c (KOI 168.01; upper left), Kepler-23b (KOI 168.03, upper right) and KOI 168.02 (lower left).  The solid curve shows the mean of the GP models as a function of time, after conditioning on the observed transit times.  The dotted lines show the 68.3\% credible interval of the GP models.  Each GP model is only affected by the TTs of one planet candidate, and yet there is a strong anti-correlation ($C_{1,3}= -0.863$) between the GP models for Kepler-23 b\&c.  In the bottom right, we show the histogram of the correlation coefficients for synthetic datasets generated by permuting the order of transit times for each of Kepler-23 b\&c, demonstrating that the observed TTVs are highly significant (FAP$<10^{-3}$).  Thus, the two bodies are in the same physical system and are not the result of two EBs or planets around two separate stars that happen to fall within the same \Kepler aperture.  The requirement of dynamical stability provides an upper limit on the masses ($\sim0.8$ \& 2.7$ M_{\mathrm{Jup}}$), allowing us to conclude that both are planets.  
}
\label{figGp168}
\label{figCHisto168}
\end{figure*}

\begin{figure*}
\epsscale{0.8}
\plottwo{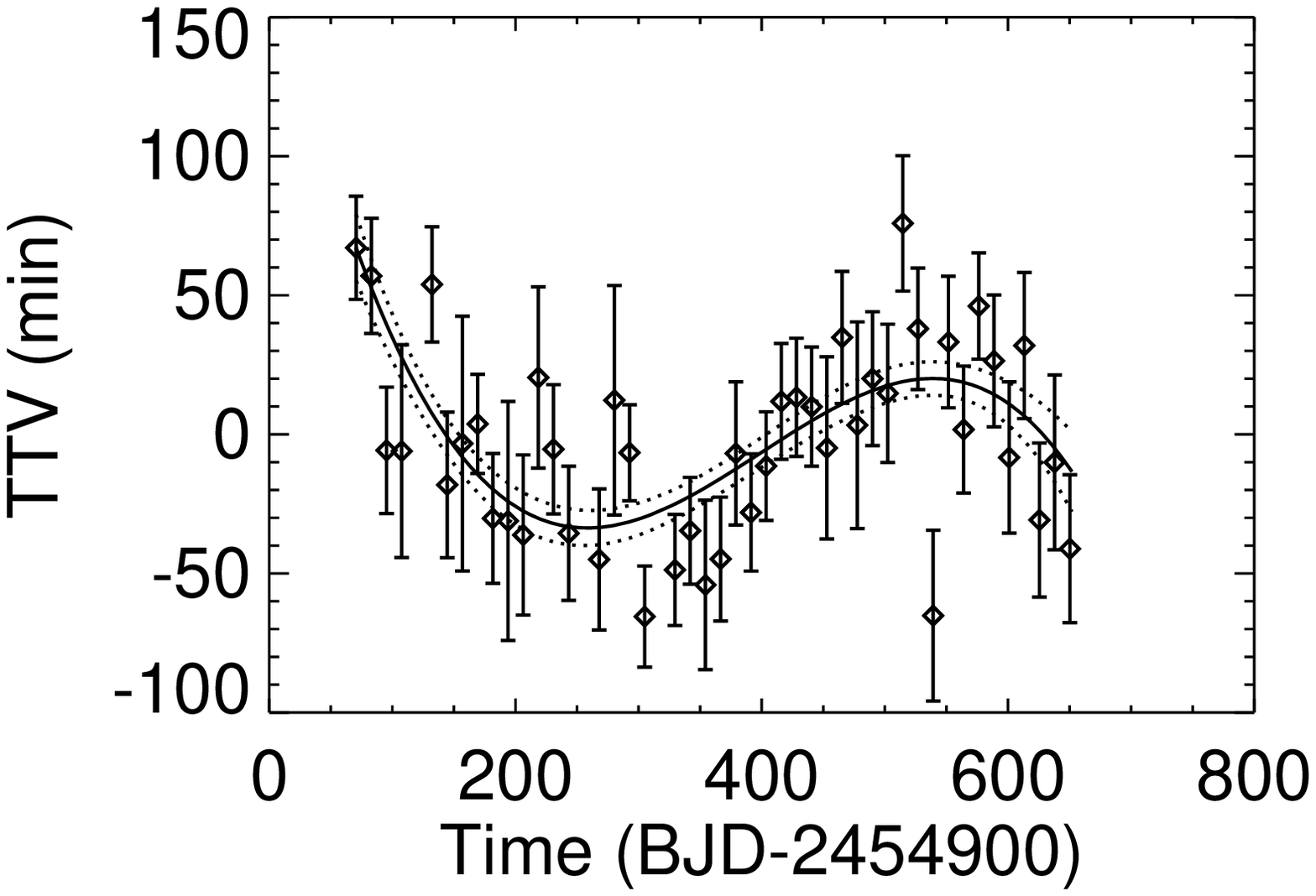}{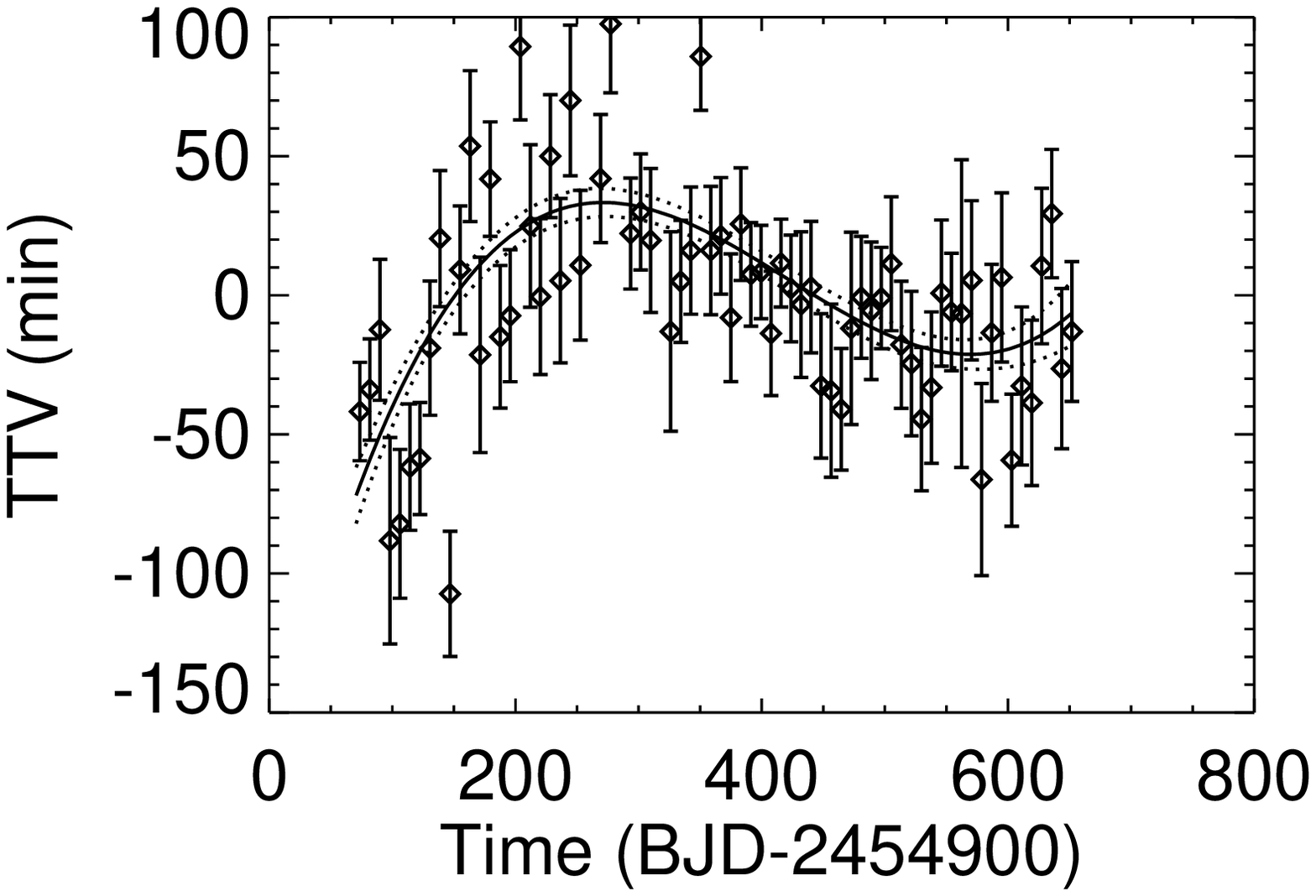}
\plottwo{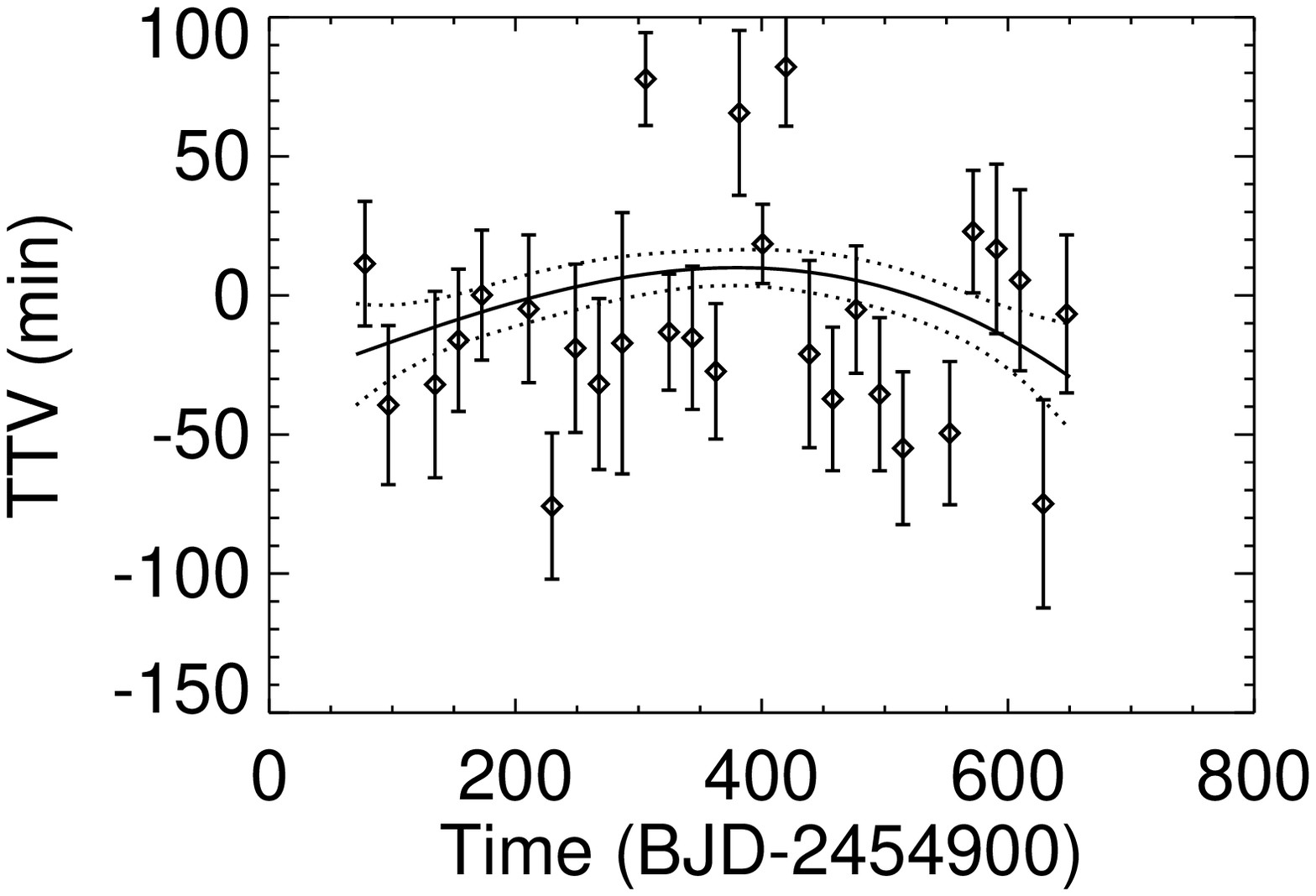}{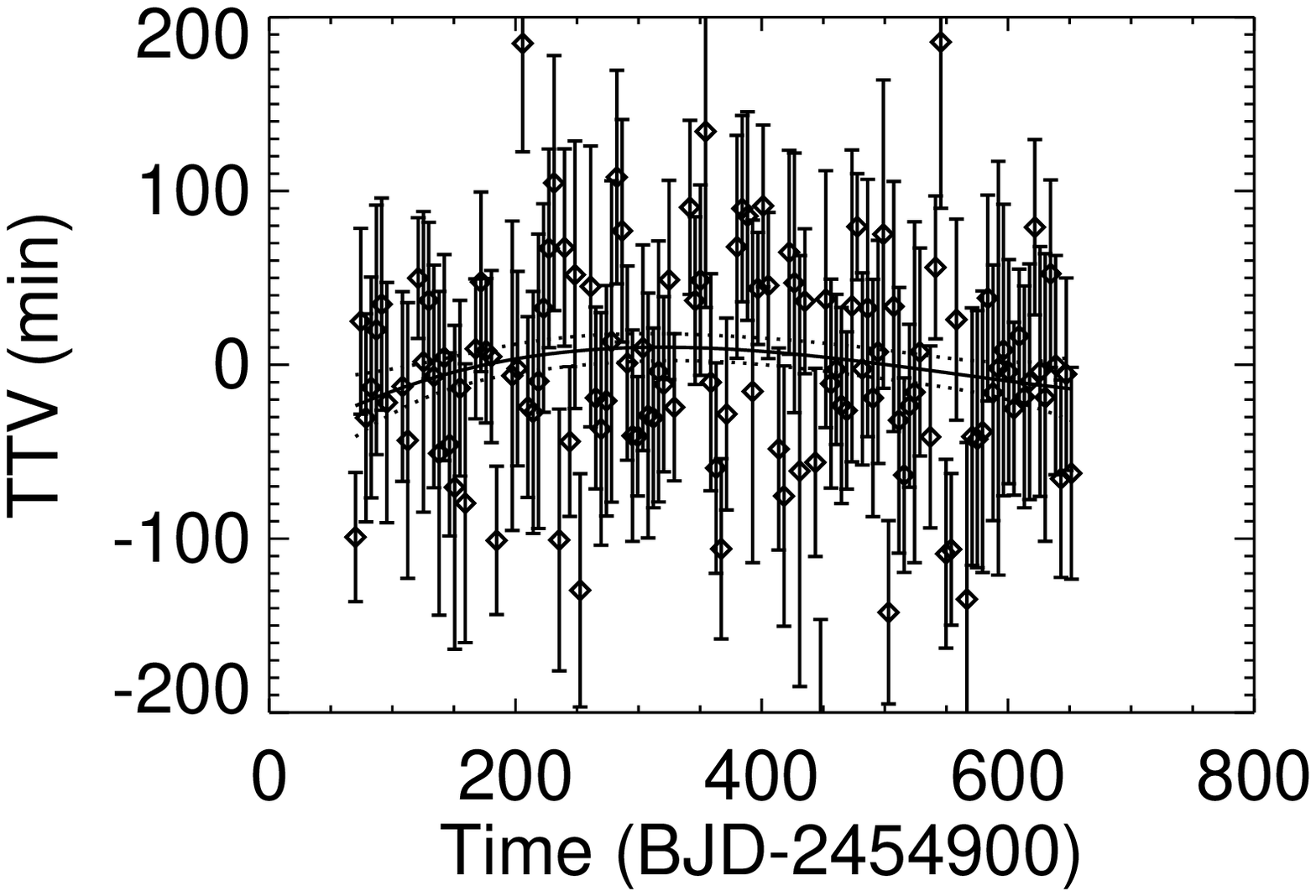}
\epsscale{0.5}
\plotone{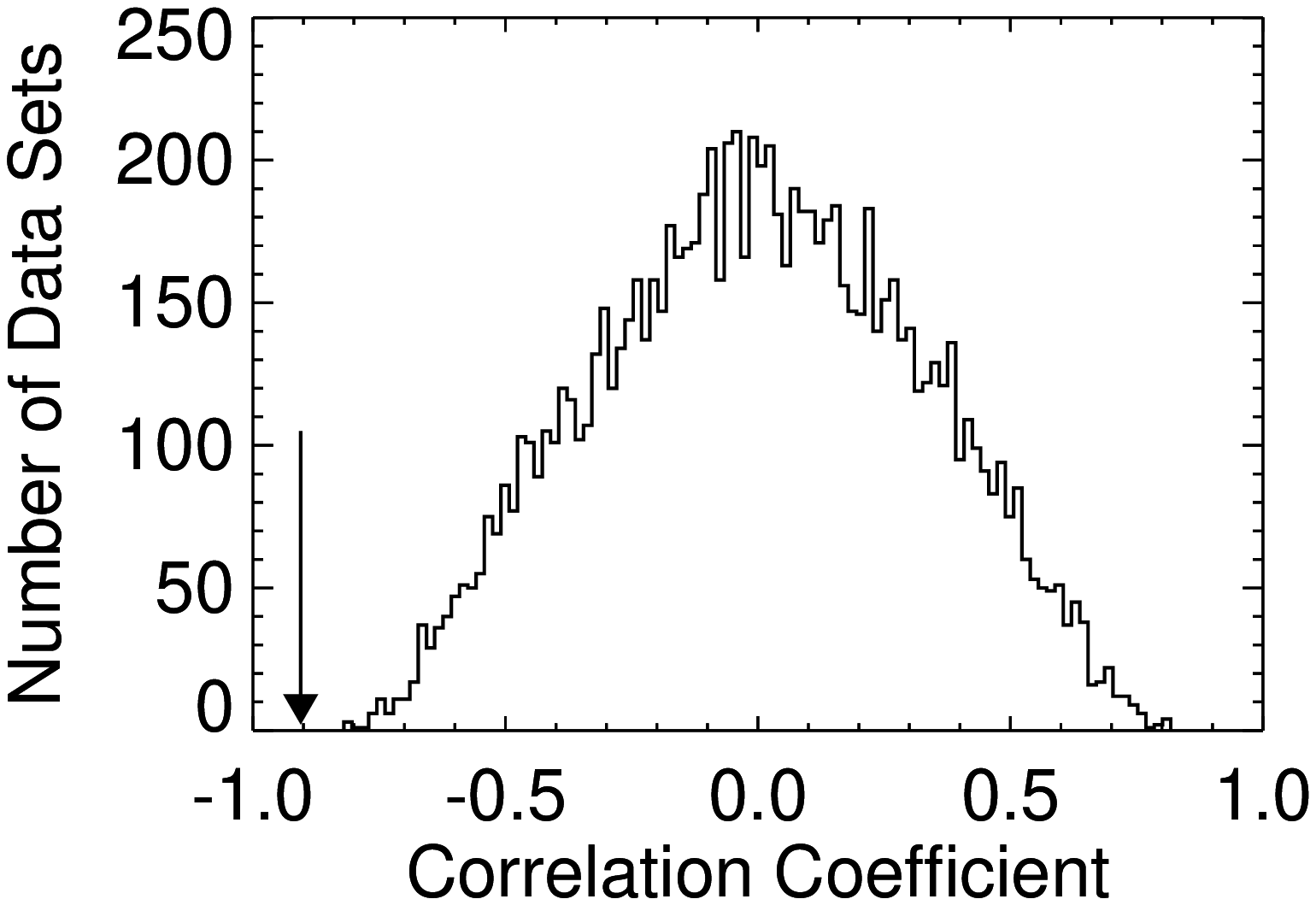}
\epsscale{1.0}
\caption{Gaussian Process models for transit times of Kepler-24's planet candidates.   
Points with error bars show deviations of measured transit times from a linear ephemeris for Kepler-24c (KOI 1102.01, upper 
left) and Kepler-24b (KOI 1102.02, upper right).  The solid curve shows the mean of the GP models as a function of time, after conditioning on the observed transit times.  The dotted lines show the 68.3\% credible interval of the GP models.  Each GP model is only affected by the TTs of one planet candidate, and yet there is a strong anti-correlation ($C_{1,2}=-0.905$) between the GP models for Kepler-24 b\&c.  At the bottom, we show the histogram of the correlation coefficients for synthetic datasets generated by permuting the order of transit times for each of Kepler-24 b\&c, demonstrating that the observed TTVs are highly significant (FAP$<10^{-3}$).  Thus, the two bodies are in the same physical system and are not the result of two EBs or planets around two separate stars that happen to fall within the same \Kepler aperture.  The requirement of dynamical stability provides an upper limit on the masses ($\sim1.6$ \& 1.6 $M_{Jup}$), allowing us to conclude that both are planets.  While there is significant uncertainty in the stellar mass, both masses would remain in the planetary regime, even if the stellar mass had been underestimated by a factor of two.  
}
\label{figGp1102}
\end{figure*}

Basic information about transit parameters, stellar and planetary properties was presented in B11.  A few key parameters are reproduced or updated in Tables \ref{tabPlanets} \& \ref{tabStars}.  
In Table \ref{tabPlanetPairs} we report the correlation coefficient ($C$) for several pairs of neighboring transiting planets, along with 
the false alarm probability (FAP$_{\rm TTV,C}$) for the TTVs based on our correlation analysis and Monte Carlo simulations (see \S\ref{secSignificance}).  The table also includes the ratio of transit durations normalized by orbital period to the one third power ($\xi$), the 5th or 95th percentile of the distribution of $\xi$ expected for a pair of planets around the target host star, the ratios of the root mean square (RMS) and mean absolution deviation (MAD) of the TTs from the best-fit linear ephemeris for the two planets.  In the final column, $\kappa$ gives the ratio of the measured MAD of TTVs for the two planets to the ratio of the predicted TTVs for the same two planets, based on our nominal n-body models.  
A subset of these systems (Kepler-23=KOI 168, Kepler-24=KOI 1102) will be discussed individually in \S\ref{secSysProp}.  
For these systems, we show the measured transit times and the GP model for a subset of these systems to be discussed in more detail in Figures \ref{figGp168}-
\ref{figGp1102}.  To help illustrate how we calculate $C_{\rm 0.001}$ and FAP$_{\rm TTV,C}$, we show the histogram of $C'$ values from synthetic data in Fig.\ \ref{figCHisto168} (bottom right panel) for Kepler-23 b\&c.  
The methods developed in this paper find significant and apparently correlated TTVs in several additional pairs of \Kepler planet candidates that are to be discussed in Cochran \etal (2011), J.-M. Desert \etal (2011), Fabrycky \etal (2011), Lissauer \etal (2011c), D. Ragozzine \etal (2012, in preparation) and Steffen \etal (2011).  \\

\subsection{Dynamical Stability Analysis \& Planet Mass Limits}
\label{secMassLimits}
The detection of significant and anti-correlated TTVs provides strong evidence that the bodies responsible for the transits are in the same physical system.  In principle, one might wonder whether the ``transits'' could actually be eclipses of stellar mass bodies.  In order to account for the TTVs, the bodies still need to be in the same physical system.  Given the similar orbital periods, any stellar companions would interact very strongly, raising serious doubts about the long-term dynamical stability of the system.  

\begin{figure*}[p]
\plotone{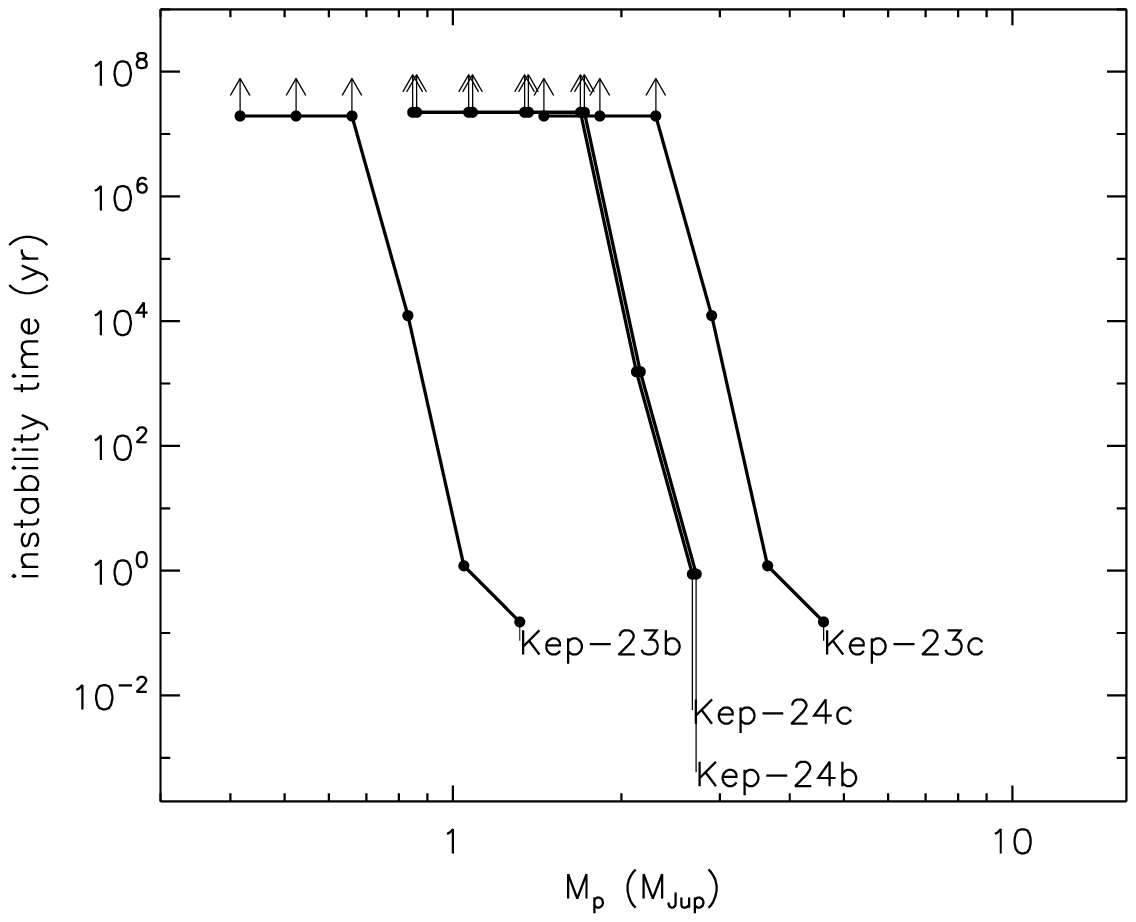} 
\caption{Timescale until dynamical instability as a function of planet mass.  For each system, we perform a series of n-body integrations including the pair of planets for which we detect significant TTVs with our correlation analysis.  We vary the masses of the planets, subject to the nominal planet-planet mass ratio based on the transit depths.  We assume initially circular, coplanar orbits and the nominal stellar masses in Table \ref{tabStars}.  Points with an upward arrow indicate n-body integrations which did not go unstable for $10^7$ years.  In all cases, the maximum masses that do not go unstable within $10^7$ years are clearly in the planetary regime.  Thus, all of the transiting planet candidates for which we observe correlated TTVs can not due to an eclipsing binary star that is blended with the target star.  
}
\label{figMassLimits}
\end{figure*}

We report the maximum planet mass for which our n-body integrations did not result in at least one body being ejected from the system or colliding with the other body or the central star (Table \ref{tabPlanets}; Fig.\ \ref{figMassLimits}).  For each of the planets presented in \S\ref{secSysProp}, the maximum mass is less than $13M_{Jup}$, excluding a triple star system as a possible false positive.   
The observed TTVs suggest even lower maximum masses, but a complete TTV analysis will require a longer time series of observations.  
There are considerable uncertainties in the stellar masses, but the available observations preclude us from having underestimated the stellar mass by more than a factor of two or more, which would be necessary for the maximum stable mass to approach $13M_{Jup}$.  
Even $\sim13M_{Jup}$ is roughly half of the recently proposed criteria for exoplanets ($\sim25M_{Jup}$; Schneider \etal 2011).  
Therefore, regardless of the choice of definition, the masses are constrained to be in the planetary regime.  

The combination of TTVs and dynamical stability provides strong evidence that the transits are due to planets orbiting a common star.  
Thus, we promote these planet candidates to confirmed planets.  
KOI 168.03 and 168.01 become Kepler-\KepNumXXX b and c, respectively.  
KOI 1102.02 and 1102.01 become Kepler-\KepNumYYY b and c, respectively.  
%
%
In \S\ref{secHostStar}, we show that the probability of the planetary system orbiting a host star other than the original target star is very small for both of the cases considered in detail.  

KOIs 168.02, 1102.03 and 1102.04 remain strong planet candidates.  
Given the low rate of false positives among the \Kepler multi-planet candidates, it is quite unlikely that these KOIs are caused by a blend with a background object.  The most likely form of a ``false positive'' would be another planetary system transiting a second physically associated and similar mass star in the \Kepler aperture.  Thus, we anticipate that further follow-up observations (such as high-resolution images) and analysis (such as BLENDER) could allow the remaining planet candidates to be validated as planets.  Alternatively, continued \Kepler TT observations may allow for dynamical confirmation of some of these candidates.

\subsection{Identification of Host Star}
\label{secHostStar}
While the correlated TTVs and dynamical stability provide evidence for a planetary system, it is not yet obvious that the system must orbit the original target star.  Here we consider the three alternate potential scenarios that could result in similar appearances:  1) both planets orbit a background star, 2) both planets orbit a significantly cooler star that is physically associated with the target star, and 3) both planets orbit one of two physically associated stars of similar mass.

The probability of the first case (planetary system around an unassociated star) can be quantified by considering the range of spectral type and magnitude differences (measured relative to the target star) which could result in a transit of the hypothetical background star mimicking the observed transit.  Since dynamical stability precludes stellar masses and an object's radius is insensitive to its mass in the Jupiter to brown dwarf-mass regime, there is a maximum size for the transiting body and the background star must be sufficiently bright that it could result in the observed transit depth (after accounting for dilution by the target star).  The maximum difference in magnitude is $\Delta$$K_{p,\mathrm max}$= 5.3 mag for Kepler-\KepNumXXX~ and $\Delta~K_{p,\mathrm max}$= 2.7 mag for Kepler-\KepNumYYY.  The potential locations for a background star are constrained by the observational limits on the centroid motion, i.e., the difference in the location of the flux centroid during transit and out of transit.  We adopt maximum angular separation equal to the $3-\sigma$ confusion radius, 0.3" for Kepler-\KepNumXXX c and 0.9" for Kepler-\KepNumYYY c.  Using a Besan\c{c}on galactic model with magnitude and position for each target star, we estimate the frequncy of background blends that could match the observed transit depth to be $\sim5\times10^{-4}$ for Kepler-\KepNumXXX~ and $\sim2\times10^{-3}$ for Kepler-\KepNumYYY.  We have not included any constraints on the observed transit duration or shape.  As \Kepler continues to observe these systems, we expect that TTVs will eventually provide significant constraints on the orbital eccentricities.  Incorporating such a constraint would be expected to rule out small host stars that would require an apocentric transit to match the observed duration.  Even if we were to assume that large planets are as common as small planets, then these planetary systems are more than $\sim10^{4}$ and $\sim5\times10^{2}$ times more likely to orbit the target star than an unassociated background star.  This excludes the vast majority of background blends, including those involving the reddest host stars.  

Next, we consider the possibility that the target star might have an undetected stellar companion that could host the planetary system.  
We can exclude blends that would result in a color $\Delta(r- K) \ge 0.1$ magnitude larger than expected based on a simple set of isochrones (Marigo \etal 2008).  Combining photometry from KIC and 2MASS, this typically rules out companions in the $\sim0.6-0.8 M_{\odot}$ range.  We will consider nearly equal mass binaries later in this section.  If the planets were to orbit a physically associated star other than the target, then the host star would be less massive than the target star and the corresponding upper mass limits for the planets would be further reduced, since the dynamical stability constraint is most closely related to the planet-star mass ratios.  For Kepler-\KepNumYYY~ the constraints based on the transit depth and dynamical stability overlap, so there are no viable blend scenarios where the planetary system orbits a physically bound and significantly lower-mass secondary star. 
For Kepler-\KepNumXXX, we compute the frequency of plausible blend scenarios using the observed frequency of stellar binaries (Raghavn \etal 2010) and the observed distribution of orbital periods and primary-secondary mass ratios (Duquennoy \& Mayor 1991).  We conservatively assume that large planets are as common as small planets and do not impose constraints based on the transit duration or shape.  
We find a blend frequency of $0.11$, indicating that the Kepler-\KepNumXXX~ system is at least $9\times$ more likely to be hosted by the primary target than by a physically bound and significantly lower-mass secondary star.  

Finally, we consider the potential for the target to be a binary star with two similar mass stars.  In many cases spectroscopic follow-up observations would have detected a second set of spectra lines.  However, we can not totally exclude a long period binary with two stars that happen to have the same radial velocity at the present epoch.  Since the two stars would have similar properties, the planet properties are largely unaffected, aside from the $\sim50\%$ dilution causing the planet radii to increase by $\sim40\%$.  As we are not overly concerned about which of two similar stars hosts the planet, this scenario essentially amounts to unseen dilution, a regular concern among faint transiting planets.

\section{Follow-up Observations \& Additional Analysis}
\label{secFop}
Some of the planet candidates investigated in this paper were not vetted in time for the results to be included in B11.  Therefore, we report the results of two tests that were instrumental in identifying many of the candidates that received a vetting flag of 3 or were labeled as likely false positives in B11.

%
We also present complementary observations obtained by the \Kepler Follow-Up Observation Program (FOP).  These results demonstrate that there are not any ``red flags'' that might indicate a more complicated system that would require a more detailed analysis.  Here we give a brief overview of the additional analysis.

\subsubsection{Odd-Even Test}
\label{secOddEven}
One of the common reasons for a KOI to have been identified as a likely false positive or to have received a vetting flag of 3 in B11 is a measurement of significant difference in the transit depth of odd and even numbered ``transits''.  This can occur if the apparent transit is due to an EB, where the odd and even ``transits'' differ in which star is eclipsing and which is being eclipsed.  (Typically, the EB must also be diluted and/or grazing in order for the depth to be consistent with a planet.)  The \Kepler pipeline provides an odd-even depth test statistic that can be used to identify KOIs potentially due to an EB (Steffen \etal 2010).  Inspecting the odd-even test statistic is also advised to check that the inferred orbital period is not half the true orbital period.  Such a misidentification can arise for low signal-to-noise candidates, such as KOI 730 (see Lissauer \etal 2011; D. Fabrycky \etal 2012, in preparation).  We have verified that the odd-even test statistic is less than 3 for each of the planets with significantly correlated TTVs that is discussed in \S\ref{secSysProp}, as well as the other planet candidates in these systems.  In the course of our analysis, we noted that the originally reported period for KOI 168.02 was an artifact at one third the period of the updated period for KOI 168.02.

\subsubsection{Centroid Motion}
\label{secCentroids}
Another of the common reasons for a KOI to receive a vetting flag of 3 in B11 was a measurement of significant difference in the location of the centroid of the target star during transit and out-of-transit.  Centroid motion can be due to to a background EB that is blended with the target star.  On the other hand, small but statistically significant centroid motion does not necessarily imply that the KOI is not a planet.  For example, local scene crowding can induce an apparent shift in the photometric centroid, as well as introduce biases in pointspread function (PSF)-fitted estimates of the location of the transiting object via difference images.
Removing these spurious sources of centroid motion requires extensive analysis as described in Bryson \etal (2012, in preparation).  Here, we limit ourselves to planet candidates where the above biases are negligible, so the apparent centroid motion is within the 3-sigma statistical error due to pixel-level noise for most of the quarters of data analyzed.  
None of the planet candidates with correlated TTVs that are discussed in \S\ref{secSysProp} have statistically significant centroid motion.

\subsubsection{Transit Durations}
\label{secDurations}
For targets with multiple transiting planet candidates, the ratio of transit durations can be used as a diagnostic to reject blend scenarios (Holman \etal 2010; Batalha \etal 2011; Lissauer \etal 2011; R. Morehead et al.\ 2012, in preparation).  
Therefore, we perform a light curve fitting based on Q0-6 data primarily to measure transit durations.  We construct folded light curves based on the measured transit times (see Table \ref{tabTTs}).  Otherwise, we follow the fitting procedure of Moorhead \etal (2011).  Note that the durations reported in Table \ref{tabPlanets} are based on when the center of the planet is coincident with the limb of the star and that this differs from the durations reported in B11 that were based on the time interval between first and fourth points of contact.  The former definition of duration is less sensitive to the uncertainties in the planet size, impact parameter and limb darkening (Col\'{o}n \& Ford 2009; Moorhead \etal 2011).  Note that the planet candidates discussed in this paper have faint host stars, so there is often a large uncertainty in the impact parameters.  
In most cases where the impact parameter is not well-constrained, we adopt a central transit, similar to both B11 and Moorhead \etal (2011).  
%
%

We report the normalized transit duration ratio ($\xi \equiv (T_{dur,in}/T_{dur,out}) (P_{out}/P_{in})^{1/3}$) for each pair of neighboring planets in Table \ref{tabPlanetPairs}.  
We also calculate $\xi_{5}$ and $\xi_{95}$, the 5th and 95th percentile of $\xi$ values obtained from Monte Carlo simulations of an ensemble of systems with two planets with the measured orbital periods and a distribution of impact parameters and eccentricities.  Simulations are discarded if both planets do not transit or if transits of one planet would not have been detected (due to grazing transits that would result in a reduced signal-to-noise ratio).  
In Table \ref{tabPlanetPairs}, we report $\xi_{5/95}$ which is simply $\xi_{5}$ for pairs with $\xi<1$ and $\xi_{95}$ for pairs with $\xi>1$.  
Typically, the distribution of $\xi$ is not far from symmetric, so $\xi_{0.95}\sim1/\xi_{0.05}$.  For a few pairs where the two planets have substantially different radii, the simulated distribution of $\xi$ is asymmetric due to the minimum signal-to-noise criterion.  
In all cases, the measured $\xi$ is consistent with a pair of planets transiting a common host star.

\subsubsection{Spectra of Host Stars}
\label{secSpectra}

The \Kepler FOP has obtained high-resolution spectra of KOI host stars from the 10m Keck I Observatory, the 3m Shane Telescope at Lick Observatory, 2.7m Harlan J. Smith Telescope at McDonald Observatory, or the 1.5m Tillinghast Reflector at Fred Lawrence Whipple Observatory (FLWO).  
The choice of observatory and exposure time were tailored to produce the desired signal to noise.  For some faint stars, an initial low-SNR reconnaissance spectra was used for initial vetting, before obtaining a second higher SNR spectrum that was used for analysis.  
For Kepler-23, stellar parameters are based on spectra from McDonald Observatory and the Nordic Optical Telescope.  Spectra were analyzed by computing the correlation function between the observed spectrum and a library of theoretical spectra using the tools described in L. Buchave \etal (2012, in preparation).  This method provides measurements of effective temperature ($T_{\rm eff}$), metallicity ([M/H]) and surface gravity ($\log(g)$), as well as a means of recognizing binary companions that would produce a second set of spectral lines. 
%
%
In the case of Kepler-24, we adopt the stellar atmospheric parameters from the KIC and reported in Borucki \etal (2011), since a high quality spectrum is not yet available.   

We report the adopted stellar atmospheric parameters in Table \ref{tabStars}.  We also update the stellar mass and radius based on Bayesian comparison to Yonsei-Yale isochrones (Yi \etal 2001).

\subsubsection{Imaging of Host Stars}
\label{secImaging}

%
%

%
%
The Kepler mission follow-up observing program includes speckle
observations obtained at the WIYN 3.5-m telescope located on Kitt Peak.
Speckle observations of Kepler-23 were used to provide high spatial resolution views of the target star to look for previously unrecognized close companions that might contaminate the \Kepler light curve.  The speckle observations make use of the Differential Speckle Survey Instrument (DSSI), a recently upgraded speckle camera described in Horch et al. (2010) and Howell et al.\ (2011). The DSSI provides simultaneous observations in two filters by employing a dichroic beam splitter and two identical EMCCDs as the imagers. 
The details of how we obtain, reduce, and analyze the speckle results and specifics about how they are used eliminate false positives and aid in transit detection are described in Torres et al.\ (2010), Horch et al.\ (2010), and Howell et al.\ (2011). The latter paper also presents the speckle imaging results for the 2010 observing season.

%
%
Classical imaging systems provide complementary observations with a wider field of view.  In particular, the Lick Observatory 1m Nickel Telescope took an I-band image of Kepler-23 with a pixel scale of 0.368"/pixel and seeing of $\sim$1.5".
%
%
For Kepler-24, the 2m Faulkes Telescope North (FTN) provides SDSS-r' band images with a pixel scale of 0.304 "/pixel in the default 2 x 2 pixel binning mode, and a typical seeing of $\sim$ 1.2". For each target a stacked image is generated by combining several images taken during the night, or on separate nights, when the target is at different positions on the sky. This is done in order to achieve increased sensitivity for faint stars without saturating the bright stars, and to average out the diffraction pattern of the spider vanes supporting the secondary mirror (Since FTN has an alt-azimuth mount the diffraction pattern is different at different sky positions, relative to the positions of the stars).  
%
%

\begin{figure*}
\includegraphics[width=3.6in]{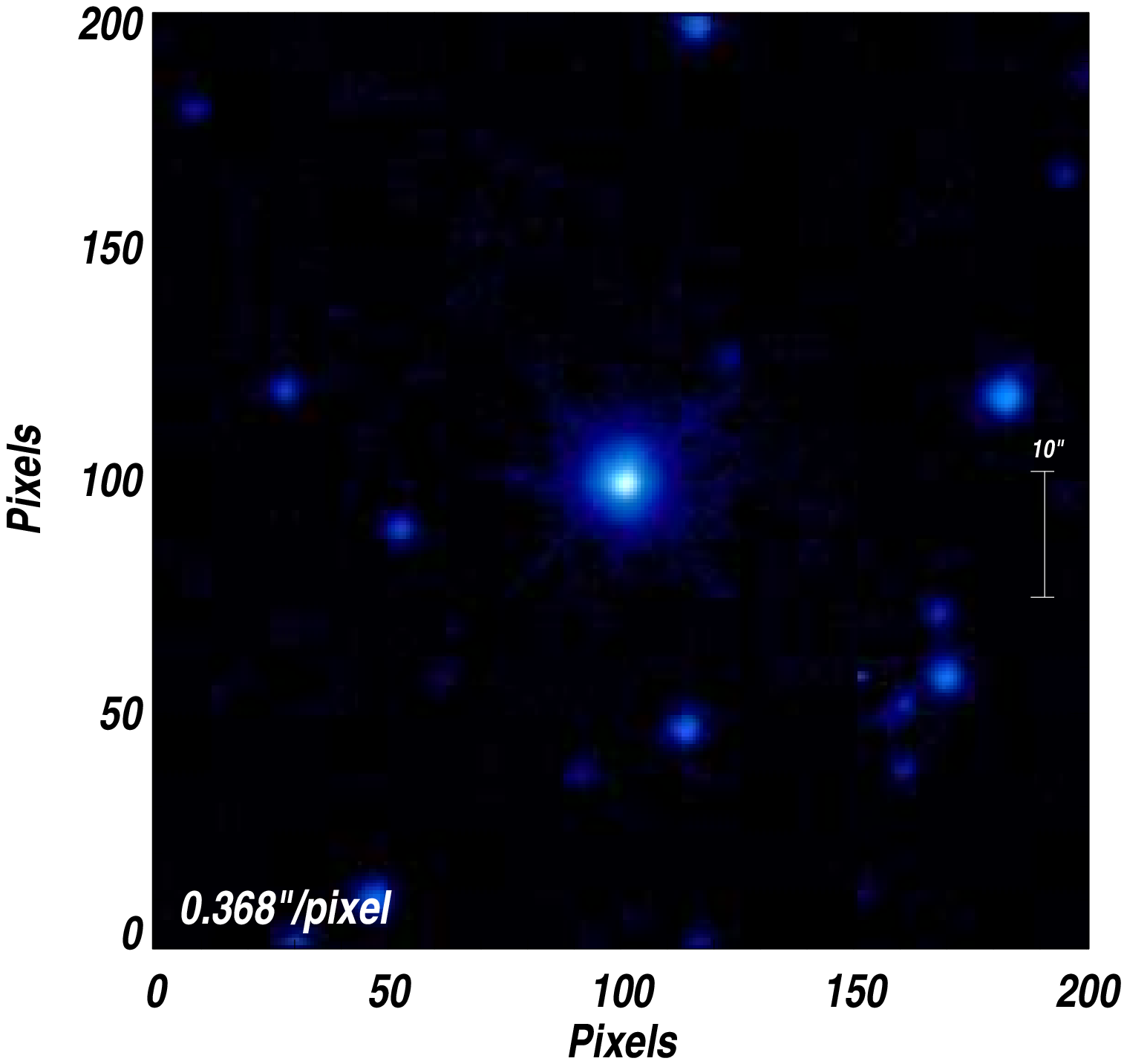}
\includegraphics[width=3.4in]{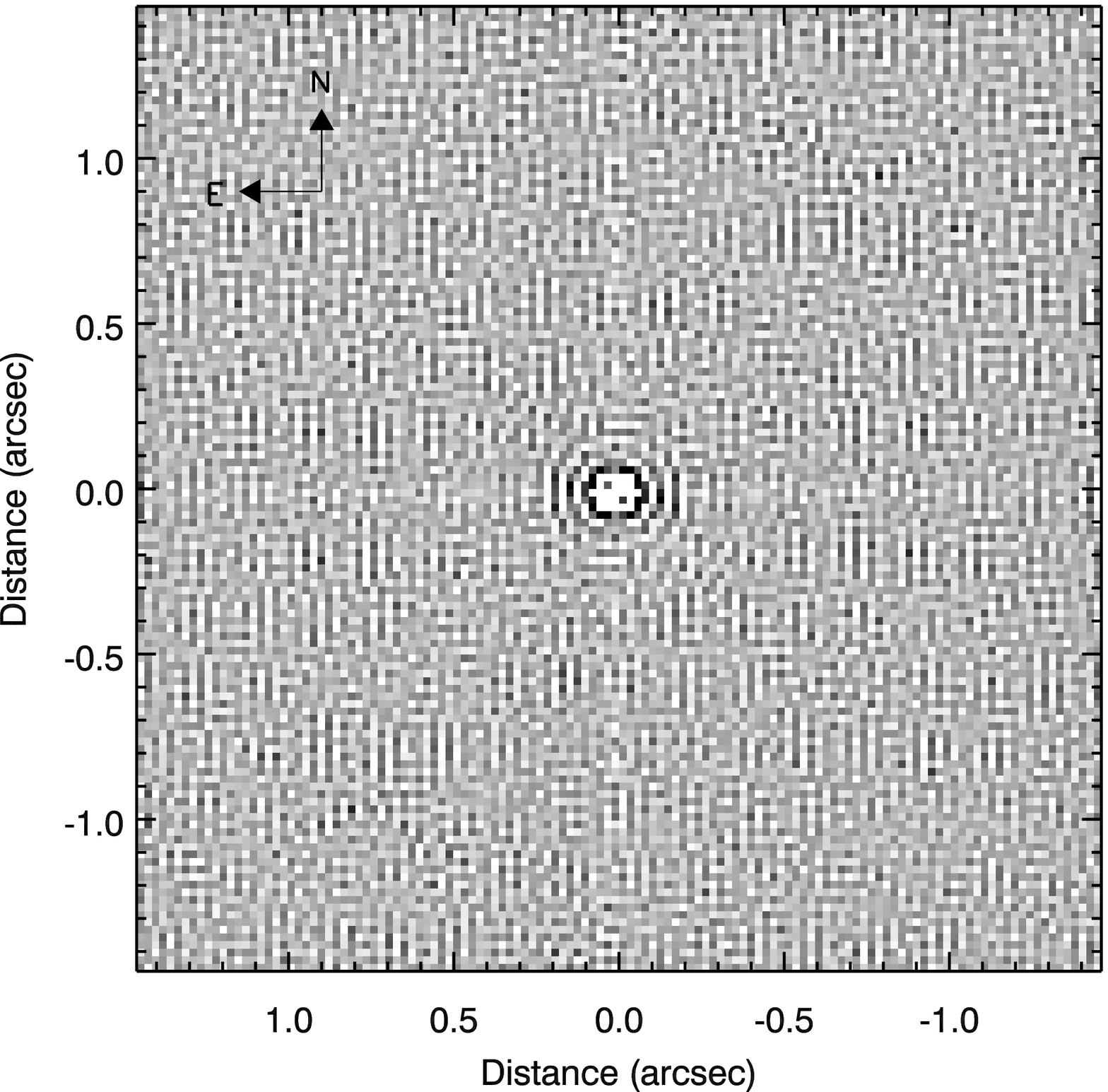} \\
\includegraphics[width=3.5in,angle=270]{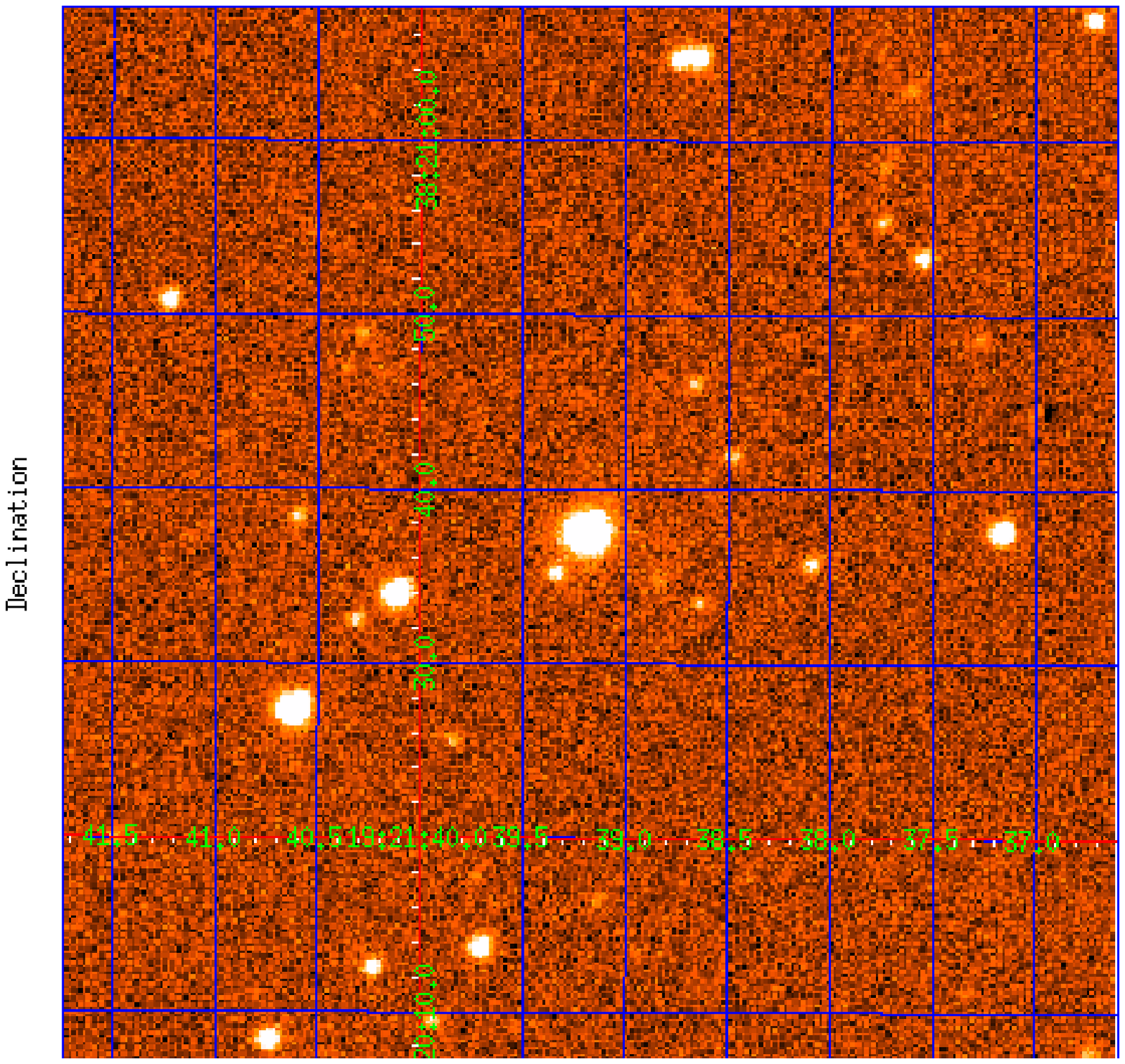}
\includegraphics[width=2.20in,angle=270]{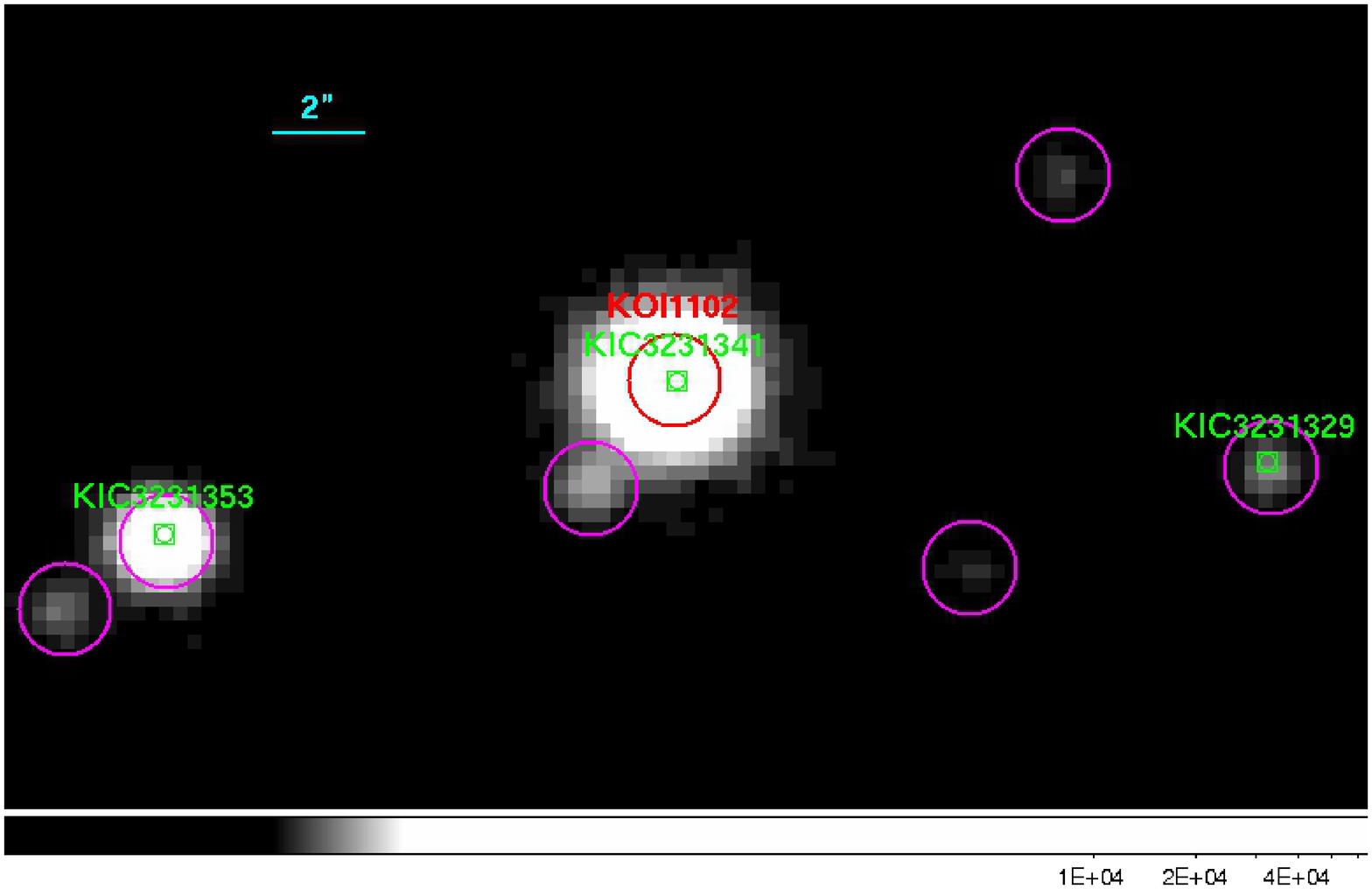}
\caption{
Upper Left:  An I-band image of Kepler-23 from the Lick Observatory 1m Nickel Telescope with 1.2' on a side.
Upper right:  A speckle image of Kepler-23 with DSSI at WIYN centered on 880nm with 2.8'' on a side.  
Lower left:  A J-band image of Kepler-24 from UKIRT, 1' on each side (North is up, East is left). 
Lower right:  A r'-band image of Kepler-24 from FTN with the target (red), KIC objects (green, with KIC ID), and other stars detected in the field (purple) circled.  Each circle has a 1" radius.
 }
\label{figImages}
\label{figImage168}
\label{figImage1102}
\end{figure*}

\section{Properties of Confirmed Planetary Systems}
\label{secSysProp}
Properties of the host stars from the \Kepler Input Catalog (KIC; Brown et al.\ 2011) and planet candidates from \Kepler light curve analysis are presented in Borucki \etal 2011.
Table \ref{tabStars} summarizes the key host star parameters, either from the KIC or the \Kepler FOP (when available).
The key properties of the planets confirmed in this paper, as well as additional planet candidates with interesting TTVs, are summarized in Table \ref{tabPlanets}. 
In this section, we discuss two planetary systems that we confirm based on correlated TTVs and dynamical stability.  Other systems with significant and strongly correlated TTVs are the subject of separate upcoming papers (Cochran \etal 2011; J.-M. Desert \etal 2011b, in preparation; D. Fabrycky \etal 2012; J. Lissauer \etal 2012; D. Ragozzine \etal 2012, in preparation; Steffen \etal 2012).

\subsection{KOI 168}
\label{secKoi168}
%
%
Kepler-23 (KOI 168, KID 11512246, $Kp$=13.4) hosts three small planet candidates (168.03, 168.01, 168.02) with orbital periods of 7.10, 10.7 and 15.3 days and radii of $\sim3.2$, 1.9 and 2.2 $R_{\oplus}$, respectively (B11; N. Batalha \etal 2012, in preparation).  In the course of this analysis, we recognized that the originally reported period for KOI 168.02 was an artifact with one third the true period.  
Initial vetting of KOI 168.01 (planet c, 10.7d) resulted in a vetting flag of 2 and an estimated EB probability of $\sim3.2\times~10^{-5}$ (B11), indicating that this candidate is very unlikely to be due to a background eclipsing binary.  KOI 168.02 and 168.03 (b) were not vetted in time for B11.  
Based on the analysis described in \S\ref{secFop}, we find no evidence pointing towards a false positive for any of the three planet candidates.
An analysis of possible centroid motion resulted in $3\sigma$ radii of confusion of $R_{c,23c}\sim0.3"$ and $R_{c,23b}\sim5"$.  While the radius of confusion for Kepler-23b is relatively large, the mean offset of the centroid during transit is only $\sim0.1\sigma$, consistent with measurement uncertainties.  Due to the faint stars and relatively low SNR of the transits, the constraint for Kepler-23b is relatively weak when compared to \Kepler planet candidates transiting brighter stars.  Nevertheless, the centroid measurements are still powerful results.  Since the typical optimal aperture for photometry of Kepler-23 is 9 pixels in area (i.e., equivalent to a circular aperture with radius of $\sim6.7$"), excluding locations for a potential background EB to within 5" eliminates 80\% of the optimal aperture phase space for blends with background objects.  The analysis in \S\ref{secHostStar} shows that the probability of a background eclipsing binary causing one of Kepler-23 b or c is less than $\sim10^{-4}$ and the probability of the planets being around a lower mass and physically bound star is less than $\sim11\%$.


An I-band image was taken with the Lick Observatory 1m Nickel Telescope on July 10, 2010.  
There were no companions visible from $\sim$2-5'' away from the target star down to 19th magnitude (see Fig.\ \ref{figImage168}).  
On Jun 11 and 12, 2011, the FOP obtained speckle images of Kepler-23 in a band centered on 880nm with width 55nm using DSSI at WIYN.  Seven and five integrations, each consisting of 1,000 40-ms exposures were coadded.  These observations reveal no secondary sources with a 4-$\sigma$ limiting delta magnitude of 2.94 at 0.2'', increasing to 3.5 at 1" and 3.57 at 1.8".  
As no new nearby stars were identified in either set of observations, we adopt $\sim2.5$\% contamination of the \Kepler aperture used by the \Kepler pipeline based on prelaunch photometry, the optimal aperture used for photometry and the PSF.

%

The \Kepler FOP obtained two medium-resolution reconisance spectra of Kepler-23 from McDonald Observatory (HJD=2455153.643746) and FIES (2455054.576815). 
The spectra result in stellar atmospheric parameters of $T_{\rm eff}=5760\pm124$K, $\log(g)=4.0\pm0.14$ and [M/H]=$-0.09\pm0.14$  (L. Buchave \etal 2012, in preparation), consistent with the stellar properties in the KIC.  We compare to the Yonei-Yale isochrones and derive values for the stellar mass ($1.11^{+0.09}_{-0.12}M_\odot$) and radius ($1.52^{+0.24}_{-0.30}R_\odot$) that are slightly smaller than those from the KIC.  We estimate a luminosity of $\sim2.3L_\odot$ and an age of $\sim4-8$Gyr.

\begin{figure*}
\epsscale{1.0}
\plotone{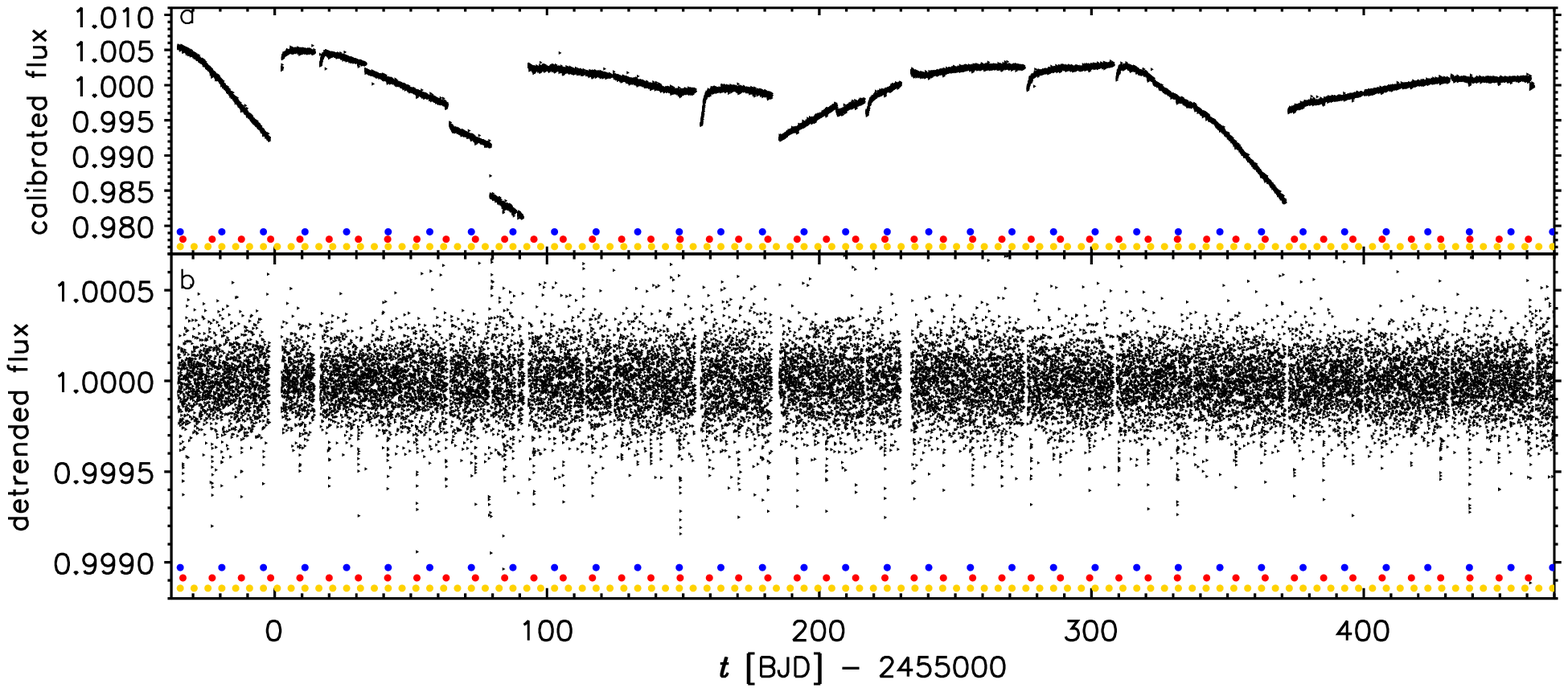} 
\epsscale{0.8}
\plotone{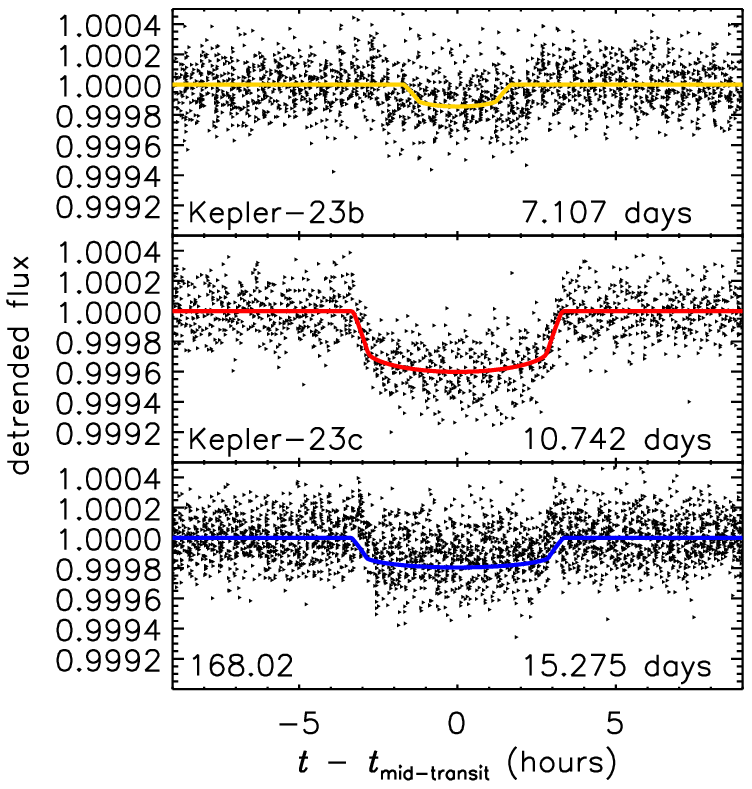}
\epsscale{1.0}
\caption{\Kepler Light curve for Kepler-23.  Panel a shows the raw calibrated \Kepler photometry (PA) and panel b shows the  photometry after detrending.  The transit times of each planet are indicated by dots at the bottom of each panel.  
The bottom, middle and top rows of dots (yellow, red, blue) are for Kepler-23 b, c and KOI 168.02.
The lower three panels show the superimposed light curve for each transit, after shifting by the measured mid-time, for Kepler-23b, c and KOI 168.02.
}
\epsscale{1.0}
\label{figLc168}
\end{figure*}

For Kepler-23b, the SNR of each transit is only $\sim1.7$, resulting in significant timing uncertainties ($\sigma_{\rm TT}=$ 47 minutes).
All three planet candidates are clearly identified after folding the light curve at the best-fit orbital period (Fig.\ \ref{figLc168}).  
%
%
Ford \etal (2011) noted that Kepler-23c was a TTV candidate based on an offset in the transit epoch between the best-fit linear ephemerides based on Q0-2 and Q0-5.  TTVs for Kepler-23b were not recognized based on Q0-2 observations alone.  

The correlation coefficient between the GP models for TTs of Kepler-23 b\&c is $C=-0.86$, well beyond the distribution of correlation coefficients calculated based on synthetic data sets with scrambled transit times (see Fig.\ \ref{figGp168}).  
The false alarm probability for such an extreme correlation coefficient is $<10^{-3}$.  
The correlation coefficient between the GP models for TTs of Kepler-23c and KOI 168.02 is -0.23 with a FAP$\sim12\%$.  This is well above our threshold for claiming a planet detection.  
We do not yet detect statistically significant TTVs for 168.02, as expected due to the smaller predicted TTV signal for KOI 168.02 and the sizable timing uncertainties.  
In combination with the analysis of \Kepler light curves and FOP observations described above, the strongly anti-correlated TTVs of Kepler-23 b \& c provide a dynamical confirmation that the two objects orbit the same star.

\begin{figure*}
\plotone{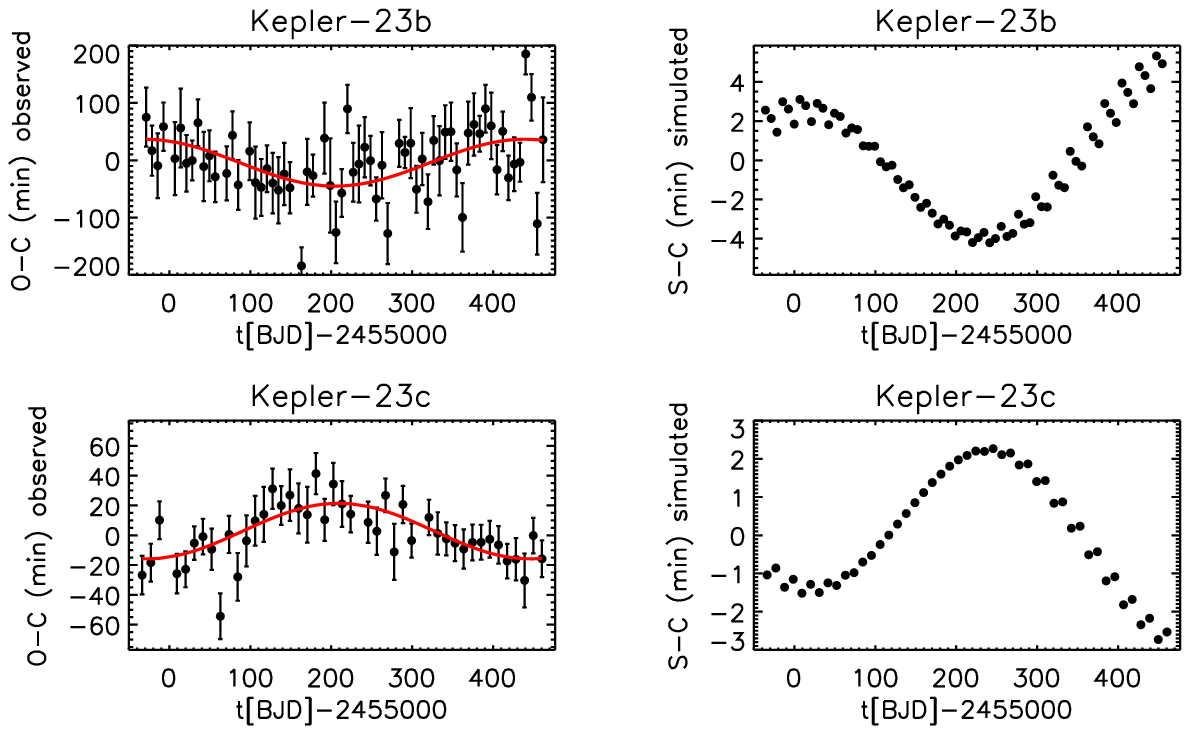}
\caption{Comparison of measured transit times (left) and transit times predicted by the nominal model (right) for a system containing only Kepler-23b (top) and c (bottom).  The model assumes circular and coplanar orbits and nominal masses, as described in \S\ref{secNbodyMethods} and Table \ref{tabStars}.  The red curves on the left show the best-fit sinusoidal model with a period constrained to be that predicted by measured orbital periods (see Fabrycky \etal 2011 for details).  The red error with no point at the far left shows the best-fit amplitude.  
}
\label{figTheory168}
\end{figure*}

Kepler-23 b\&c have short orbital periods and lie close to a 3:2 resonance: $P_{23b}/P_{23c}=1.511$.  Therefore, we expect a relatively short TTV timescale:
\begin{equation}
P_{\mathrm TTV} = 1 / (3/ 10.7421 d - 2 / 7.1073 d) = 470 d.
\end{equation}
The observed timescale ($\sim455$d) and phase of TTVs for Kepler-23 b\&c are consistent with expectations based on n-body integrations using nominal planet mass estimates and circular, coplanar orbits (see Fig.\ \ref{figTheory168}).  The amplitude of the observed TTVs is $\sim5$ times larger than the nominal model.  Such differences are not unexpected, as even modest eccentricities can significantly increase the TTV amplitude (e.g., Veras \etal 2011).  The prediction of the nominal model for the ratio of the TTV amplitudes of the two confirmed planets is significantly more robust than the predictions of individual amplitudes.  Indeed, the observed MAD TTV from a linear ephemeris is within $5\%$ of that predicted by the nominal model.  
Our GP models for the TTs of Kepler-23 b\&c are non-parametric, so there is nothing in our model that would cause the GPs to match the ratio of TTV amplitudes, the TTV period, or the phase of TTV variations.  The fact that our  general non-parametric model naturally reproduces these features provides an even more compelling case for the confirmation of these planets via TTVs.  

While the currently available data is insufficient for robust determinations of the planet masses, we fit two n-body models to the TTV observations.  If we assume initially circular and coplanar orbits, then the best-fit masses are $\sim22\pm~6$ and $12\pm2 M_\oplus$.  If we assume eccentric coplanar orbits, then the best-fit masses are $\sim15$ and 5$M_\oplus$, corresponding to orbital solutions with low eccentricities.  Significantly smaller or larger masses are possible, but these require large eccentricities.  Either model represents a significant improvement in the quality of the fit relative to a non-interacting model.  
Despite the sizable uncertainties in the estimates for the planet masses, the requirement of dynamical stability provides firm upper limits (2.7 \& 0.8 $M_{\rm Jup}$) on the planet masses (Fig.\ \ref{figMassLimits}).  

\subsection{Kepler-24 (KOI 1102)}
\label{secKoi1102}
%
%
Kepler-24 (KOI 1102, KID 3231341, $Kp$=14.9) hosts four planet candidates (04, 02, 01 and 03) with orbital periods of 4.2, 8.1, 12.3, 19.0 days and radii of 1.7, 2.4, 2.8, 1.7 $R_{\oplus}$, respectively (see Fig.\ 7; B11; N. Batalha \etal 2012 in preparation).  
KOI 1102.02 (b) and 1102.01 (c) were identified, but not vetted in time for B11.  Based on the analysis described in \S\ref{secFop}, we find no evidence pointing towards a false positive for any of KOI 1102.01-1102.04.  
The centroids during transit for Kepler-24b \& c differ from those out-of-transit by $\sim$1.4 and 2.8$\sigma$, consistent with measurement uncertainties, particularly when considering the actual distribution of the apparent centroid motion during transit is highly non-Gaussian for other well-studied cases (Batalha \etal 2010).  The $\sim3\sigma$ radii of confusion are 1.2" and 0.9", for Kepler-24 b \& c, respectively.  
Based on a preliminary analysis of data through Q8, the centroid offsets for 1102.04 and 1102.03 are less than $3\sigma$.  Again, the limited centroid motion enables us to exclude the locations for potential background EBs to a small fraction of the optimal aperture phase space for blends with background objects.  
The analysis in \S\ref{secHostStar} shows that the probability of a background eclipsing binary causing one of KOI 1102.01 (c) or 1102.02 (b) is less than $\sim2\times10^{-3}$.  The only acceptable blend scenario involves a pair of similar mass and physically bound stars.


A UKIRT J-band image (Fig.\ \ref{figImage1102}, left) reveals a faint companion that is not in the KIC $\sim2.5"$ to the SE, with brightness intermediate to the outlying faint stars KIC 3231329 ($Kp$= 19.0) and KIC 3231353 ($Kp$= 20.0).  An FTN image also reveals the nearby star, 
$\sim5$ magnitudes fainter than the target in the r'-band (Fig.\ \ref{figImage1102}, right).  We estimate that it is $\sim4-5$ magnitudes fainter than Kepler-24 in the \Kepler band.  As most of its light falls in the \Kepler image and star is not in the KIC, it results in $\sim1-3\%$ dilution in addition to the $\sim8\%$ dilution estimated by the pipeline for a total of 8.6\%.
%


The \Kepler FOP has not yet obtained a spectrum of Kepler-24.  We estimate the stellar properties based on multi-band photometry from the Kepler Input Catalog:  $T_{\rm eff}=5800\pm~200$, $\log(g)=4.34\pm~0.3$ and [M/H]=$-0.24\pm~0.40$.  We estimate uncertainties based on a comparison of KIC and spectroscopic parameters for other \Kepler targets (Brown \etal 2011).  We derive values for the stellar mass ($1.03^{+0.11}_{-0.14}M_\odot$), radius ($1.07^{+0.16}_{-0.23}R_\odot$), luminosity $\sim1.16^{+0.36}_{-0.60}~L_\odot$ and age $\le~7.7$Gyr by comparing to the Yonei-Yale isochrones.  We caution that there are significant uncertainties associated with the stellar models and derived properties.


Kepler-24 is sufficiently faint that transit times for Kepler-24c and Kepler-24b have sizable uncertainties ($\sigma_{\rm TT}= 30, 25$ min).  
Ford \etal (2011) did not find significant evidence for TTVs in either of the planet candidates, but noted that Kepler-24b was likely to have significant TTVs ($\sim26$ min), based on numerical integrations of two planet systems with nominal masses and circular orbits.  

The correlation coefficient between the GP models for TTs of Kepler-24b \& c is $C=-0.905$, well beyond the distribution of correlation coefficients calculated based on synthetic data sets with scrambled transit times (see Fig.\ \ref{figGp1102}).  
The false alarm probability for such an extreme correlation coefficient is $<10^{-3}$.  
In combination with the analysis of \Kepler light curves described above, this provides a dynamical confirmation that the two objects orbit the same star.  

%
Kepler-24b \& c have short orbital periods and lie near a 3:2 resonance: $P_{24c}/P_{24b}=1.514$.  Therefore, we expect a relatively short TTV timescale:
\begin{equation}
P_{ttv} = 1 / (2 / 8.1451 d - 3/ 12.3336 d) = 421 d.
\end{equation}
The observed timescale ($\sim400$d) and phase of TTVs for Kepler-24 b \& c are consistent with the expectations based on n-body integrations using nominal planet mass estimates and circular, coplanar orbits (see Fig.\ \ref{figTheory1102}).  Our GP models for the TTs of Kepler-24 b \& c are non-parametric, so there is nothing in our model that would cause the GPs to match the period and phase of the TTV variations.  The fact that our general non-parametric model naturally reproduces these features provides an even more compelling case for the confirmation of these planets via TTVs. 

The best-fit amplitudes of the observed TTVs are $\sim7$ and 9 times larger than the nominal model for Kepler-24b \& c, respectively.
Such differences are not unexpected, as even modest eccentricities can significantly increase the TTV amplitude (e.g., Veras \etal 2011).  
Like Kepler-\KepNumXXX, the prediction of the nominal model for the ratio of the TTV amplitudes of the two confirmed planets is within 25\% of the observed ratio (1.25). 
Still, there could be significant differences between the true masses and the nominal masses, e.g., if there are significant eccentricities and/or perturbations from other planets in the system.  Assuming that KOI 1102.03 is indeed a planet in the same system, then Kepler-24c would be near a 3:2 MMR with both Kepler-24b (inside) and KOI 1102.03 (outside), leading to complex dynamical interactions.  

While the currently available data is insufficient for robust determinations of the planet masses, we fit two n-body models to the TTV observations.  If we assume initially circular and coplanar orbits, then the best-fit masses are $\sim17\pm~4$ and $5\pm~3 M_\oplus$.  If we assume eccentric coplanar orbits, then the best-fit masses are $\sim102\pm~21$ and $56\pm~16 M_\oplus$, corresponding to orbital solutions with eccentricities of $\sim0.2-0.4$.  Either model represents a significant improvement in the quality of the fit relative to a non-interacting model.  While the best-fit eccentric model represents a significant improvement in the goodness-of-fit relative to the best-fit circular model, such large eccentricities may complicate long-term stability, particularly if all four of the planet candidates were confirmed.  
Despite the sizable uncertainties in the estimates for the planet masses, the requirement of dynamical stability provides a firm upper limit on the planet masses (1.6 \& $1.6 M_{\rm Jup}$; Fig.\ \ref{figMassLimits}).

We do not detect a statistically significant correlation coefficient between TTVs of either KOI 1102.03 or 1102.04 and any of the other planet candidates.  The scatter of the TTVs for 1102.04 is consistent with the measurement uncertainties. One can understand the lack of a TTV detection for KOI 1102.04 analytically, since the amplitude of a TTV signal scales with the orbital period, KOI 1102.04 has an orbital period roughly half that of Kepler-24b, and the TTV signal for Kepler-24b is near the limit of detectability. Further, the transit of KOI 1102.04 is shallower than Kepler-24b, so the TTVs are less precise. 
If KOI 1102.04 is a planet in the same system, then the current innermost period ratio would be $P_{24b}/P_{1102.04}=1.92$, $\sim5\%$ less than 2, whereas it is more common for systems near a 1:2 MMR to have a period ratio $\sim5\%$ greater than 2 (Lissauer \etal 2011b).  This could be related to the presence of four planet candidates with orbital period ratios near a 2:4:6:9 resonant chain.  
The current period ratios for the outer two pairs ($P_{24c}/P_{24b}=1.514$ and $P_{1102.03}/P_{24c}=1.54$) are slightly greater than 1.5, so none of the neighboring pairs are necessarily ``in resonance''.  Such deviations from an exact period commensurability appear to be common among \Kepler planet candidates (Lissauer \etal 2011).  

There may be early hints of excess scatter in the TTVs of 1102.03, but this is not statistically significant based on the current data set.  
If KOI 1102.03 is a planet in the same system, then Kepler-24c would feel significant perturbations from both Kepler-24b (on the inside) and 1102.03 (on the outside).  This may help explain why the observed TTVs for Kepler-24b are significantly greater than those predicted by the nominal two-planet model that includes only planets b \& c.  We hope that this paper will motivate more detailed analysis of the TTVs and orbital dynamics of this fascinating planetary system.

\begin{figure*}
\epsscale{1.0}
\plotone{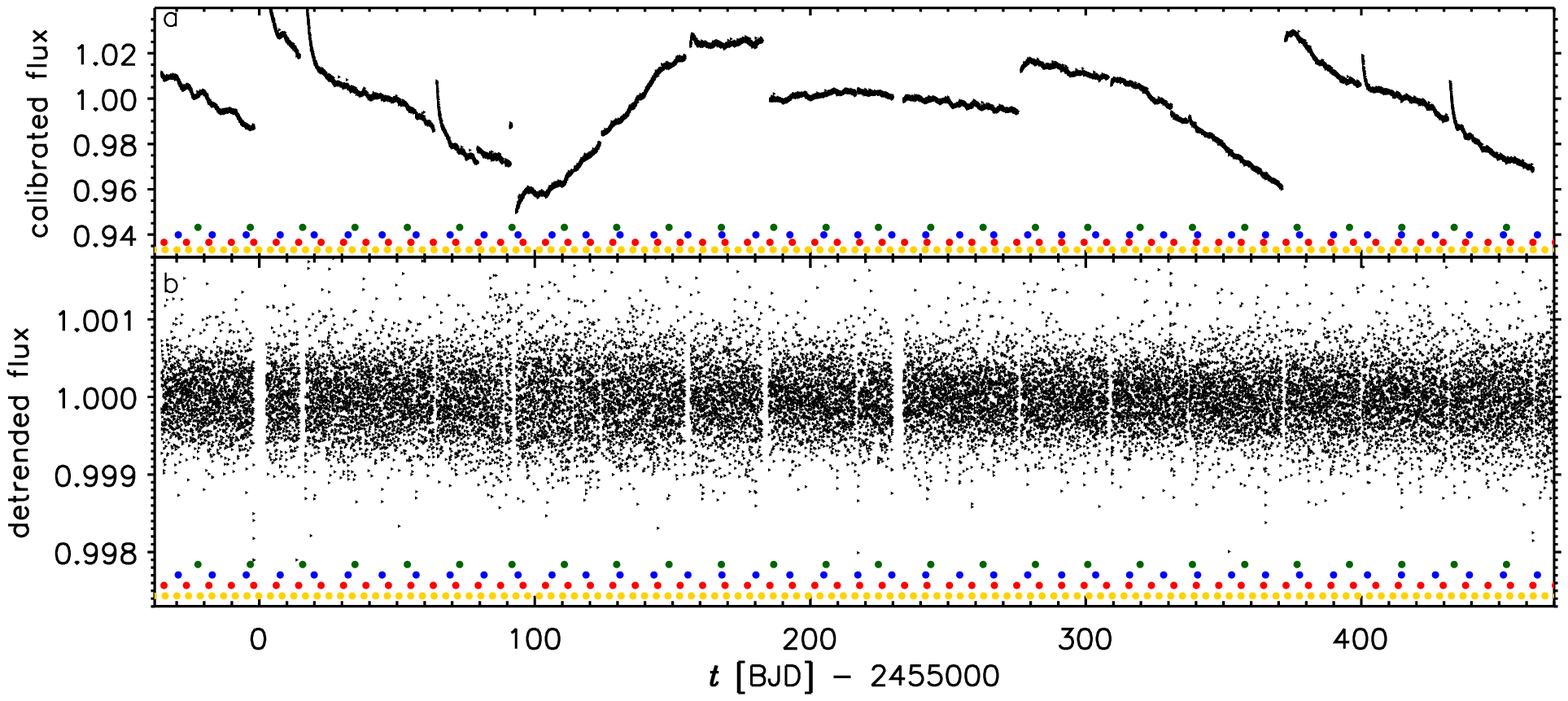} 
\epsscale{0.8}
\plotone{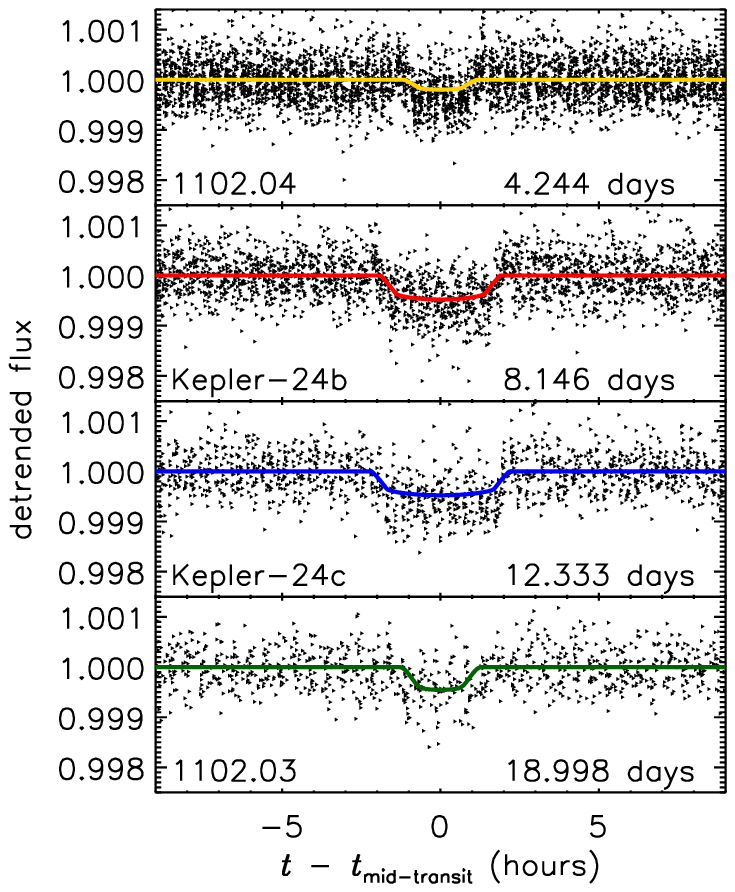}
\epsscale{1.0}
\caption{\Kepler Light curve for Kepler-24.  Panel a shows the raw calibrated \Kepler photometry (PA) and panel b shows the  photometry after detrending.  The transit times of each planet are indicated by dots at the bottom of each panel.  
From from top to bottom, the rows of dots (yellow, red, blue, green) are for KOI 1102.04, Kepler-24b, Kepler-24c and KOI 1102.03.
The lower four panels show the superimposed light curve for each transit, after shifting by the measured mid-time for each of the four KOIs.
}
\epsscale{1.0}
\label{figLc1102}
\end{figure*}

\begin{figure*}
\plotone{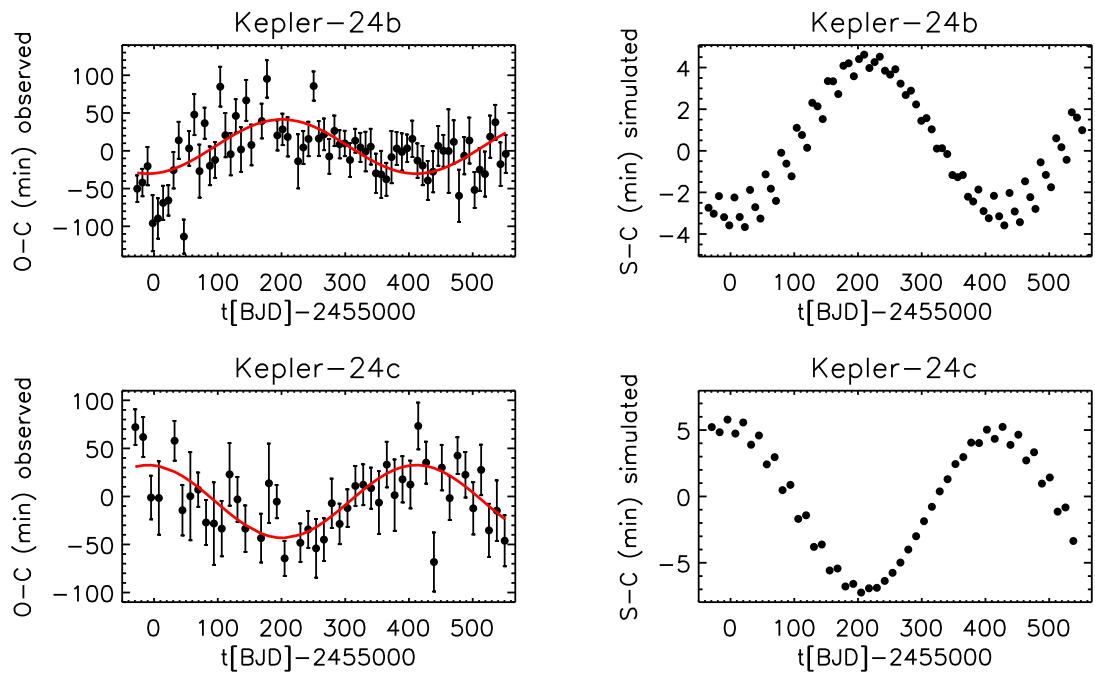} \\
\caption{Comparison of measured transit times (left) and transit times predicted by the nominal model (right) for a system containing only Kepler-24b (top) and Kepler-24c (bottom).  Details are described in the caption to Fig.\ \ref{figTheory168}.
}
\label{figTheory1102}
\end{figure*}

\clearpage

\section{Discussion}
\label{secDiscuss}
In this paper and two companion papers (Fabrycky \etal 2011, Steffen \etal 2011), we have described a new approach to confirming transiting planets.  For systems with MTPCs, correlated TTVs provide strong evidence that both transiting objects are in the same system.  Dynamical stability provides an upper limit on the masses of the transiting bodies.  For closely-spaced pairs, the mass upper limit is often in the planetary regime, allowing planets to be confirmed by the combination of correlated TTVs and the constraint of dynamical stability.  

We have described a non-parametric method for quantifying the significance of TTVs among MTPC systems.  Our approach uses a GP model to allow for a rigorous, yet computationally tractable, Bayesian analysis of each planet candidate's TTV curve.  Provided there is a sufficient number of transit observations, our correlation-based analysis is more sensitive and robust than testing for TTVs of each planet individually (e.g., Ford \etal 2011).  N-body simulations show that TTVs of two interacting planets are correlated, regardless of the exact mechanims responsible for the TTVs (e.g., resonant interaction, indirect effect on the star, secular precession).  Thus, our algorithm is designed to be most sensitive for detecting TTVs when applied to systems with correlated TTVs.  Another advantage of this method is that it makes minimal assumptions about the potential TTV signatures.  
This approach is complementary to other methods of quantifying the significance of TTVs in MTPC systems (Fabrycky \etal 2011; Steffen \etal 2011).  Since the method of Steffen \etal (2011) assumes sinusoidal TTV signals, it is expected to be more sensitive to systems with sinusoidal TTVs, but less sensitive to systems with more complex TTV signatures.  The method of Fabrycky \etal (2011) assumes that the TTVs can be approximated as sinusoidal and that the dominant TTV timescale can be predicted based on the period of a known, transiting planet.  Since this method performs a minimal number of statistical tests, it is expected to have an even greater sensitivity for many systems.  Of course, it could overlook systems where the TTV signal is more complex or dominated by the perturbations of a non-transiting planet.  Therefore, it will be useful to apply all three methods for detecting TTVs to the \Kepler MTPC systems.  We anticipate that the algorithm described here may be particularly useful for confirming closely-spaced systems or systems with more than two planets contributing to the TTV signature.

We have applied our method to \Nsystab~MTPC systems to calculate the correlation coefficient between TTV curves for neighboring transiting planet candidates.  Eight of these systems have at least one pair of neighboring planets with correlated TTVs for which the TTVs are significant at better than the $10^{-3}$ level (see Table \ref{tabPlanetPairs}).  This demonstrates that the planet candidates are in the same physical system.  Given their close proximity, the requirement of dynamical stability provides limits on the maximum masses that are firmly in the planetary regime.  This combination of correlated TTVs plus dynamical stability provides dynamical confirmation of planet candidates originally identified photometrically by the \Kepler mission.  

While correlated TTVs clearly demonstrate that the planets orbit the same star, it is not guaranteed that the host star is the target.  Based on follow-up imaging and limits on the extent of centroid motion during transit, blends with background objects are extremely unlikely.  However, it is possible that both planets orbit a physically associated star that is too close to the primary target star to be identified by imaging, but far enough away that it does not destabilize the system.  If the host star were significantly less massive and less luminous than the primary target star, then the planets would have smaller masses, but be much larger in size, since the transit depths would be highly diluted by the primary target star.  While this is not a viable scenario for Kepler-\KepNumYYY, we estimate there is a $\le11\%$ chance that the Kepler-\KepNumXXX~planetary system orbits a significantly lower mass star that is physically bound to the primary target star.  Alternatively, if the secondary is nearly the same mass and luminosity as the primary target star, then the stellar properties would be quite similar, but the planet sizes could be larger by a factor of $\sim\sqrt{2}$, due to the extra dilution.  The probability for such scenarios is small, but not completely negligible (e.g., a few percent for KOI 738; Fabrycky \etal 2011).  In any case, the correlated TTVs and the mass limits from dynamical stability still demonstrate that there are at least two planets in the same system, despite the uncertainty in stellar parameters.

This paper presents \Nsysdisc~planetary systems (Kepler-\KepNumXXX~ \& Kepler-\KepNumYYY) containing \Npldisc~confirmed planets.  
Both pairs (Kepler-\KepNumXXX b(03) \& c(01); Kepler-\KepNumYYY b(02) \& c(01) ) of confirmed planets are near the 3:2 MMR.  Each of these systems contains additional transiting planet candidates which have yet to be confirmed.  If confirmed, these systems would contain closely-spaced near-resonant chains of transiting planets (4:6:9 for Kepler-\KepNumXXX~ and 2:4:6:9 for Kepler-\KepNumYYY).  Resonant chains can form via migration through a protoplanetary disk (e.g., Cresswell \& Nelson 2006, 2008), but forming chains of near-resonant planets may be more challenging.  
More detailed modeling of the TTV curve will become a powerful tool for learning about these systems and planet formation and migration (Ragozzine \& Holman 2010), particularly once multiple cycles of TTVs have been measured by {\sl Kepler}.  


At the moment, the time span of observations is not yet sufficient to enable precise mass measurements of planets in these systems due to degeneracies with the other orbital properties.  Nevertheless, we can already recognize one of the timescales of the observed TTVs for each system.  For each of the confirmed planets, this timescale is consistent with the predictions of a nominal model, using coplanar, circular orbits and planet masses crudely estimated from their radii and the host star properties.  Additionally, the phases of the TTV curves for Kepler-\KepNumXXX b \& c and Kepler-\KepNumYYY b\&c  are consistent with the predictions of the nominal model.  
%

The observed TTV amplitudes (measured in terms of the mean absolute deviation from a linear ephemeris) are larger (factors of $\sim5.3$ for Kepler-\KepNumXXX, and  
$\sim$5-46 for Kepler-\KepNumYYY) than the predictions of our nominal models.  
Again, this is not unexpected due to the extreme sensitivity of TTVs to masses and eccentricities.  The ratio of the TTV amplitudes for each pair of planets with correlated TTVs is a more robust prediction, as several sources of uncertainty affect both planet masses similarly (i.e., stellar mass, amount of dilution of \Kepler light curve).  Indeed, we find excellent agreement ($\sim1-2\%$) between the ratio of observed TTV amplitudes and with the corresponding prediction of the nominal model for Kepler-\KepNumXXX b \& c.
For Kepler-\KepNumYYY, this ratio is $\sim0.1$, perhaps due to significant eccentricities or perturbations from KOI 1102.03.  

We caution that the uncertainty in the masses and sizes of the host stars directly translates into uncertainties in the planet masses and sizes.   
%
As the time span of \Kepler TTV observations increases, more detailed dynamical analyses will become possible.  Additional follow-up observations and analysis will become increasingly important to aid in the interpretation of the detailed dynamical information contained in the TTV curves for these systems.

Alternative techniques for studying TTVs of MTPC systems also confirm the TTVs of Kepler-\KepNumXXX~ and Kepler-\KepNumYYY~ with a false alarm probability of $<10^{-3}$ (Fabrycky \etal 2011; Steffen \etal 2011).  
%
%
The GP method here is notable for its minimal set of assumptions, making it most sensitive for large data sets, even if the observed TTV signature is complex and differs from that expected.  Thus, the GP method could prove particularly valuable for analyzing TTVs of planets that are significantly perturbed by a non-transiting planet and/or multiple planets.  

The planets confirmed here and in companion papers (Fabrycky \etal 2011; Steffen \etal 2011) are not meant to represent an exhaustive search of the \Kepler planet candidates presented in B11.  
With continued \Kepler observations, this and other complementary techniques (e.g., Fabrycky \etal 2011; Lissauer \etal 2011bc; Steffen \etal 2011) are poised to confirm many more MTPC systems.
In particular, there are several other MTPC systems near the 3:2 MMR, 2:1 MMR or other period commensurabilities that are predicted to have observable TTV signatures (Ford \etal 2011; Lissauer \etal 2011).  
For many of these KOIs additional analysis and/or observations will be required before their planets can be confirmed.  

  
%
Due to larger stellar activity than originally anticipated (Gilliland \etal 2011), for {\em Kepler} to measure the frequency of Earth-size planets in the habitable zone of solar-type stars an extended mission is  required.  Fortunately, the spacecraft carries consumables that could support an extended mission which would improve sensitivity for detecting small planets.  An extended mission would also dramatically improve the constraints on planet masses and orbits based on TTVs for systems such as those presented here.  
Indeed, of the four planets confirmed here, only one was recognized as a TTV candidate in Ford \etal (2011).  As the time span of observations increases, we anticipate that TTVs will become detectable for even more systems.  
An extended mission could also substantially increase the number of transits observed for planets at longer orbital periods.  This would be particularly valuable for confirming small planets in or near the habitable zone.

\acknowledgements  Funding for this mission is provided by NASA's Science Mission Directorate.  We thank the entire Kepler team for the many years of work that is proving so successful.  
E.B.F acknowledges support by the National Aeronautics and Space Administration under grant NNX08AR04G issued through the Kepler Participating Scientist Program.  This material is based upon work supported by the National Science Foundation under Grant No. 0707203.
D. C. F. and J. A. C. acknowledge support for this work was provided by NASA through Hubble Fellowship grants \#HF-51272.01-A and \#HF-51267.01-A awarded by the Space Telescope Science Institute, which is operated by the Association of Universities for Research in Astronomy, Inc., for NASA, under contract NAS 5-26555.
Results are based in part on observations obtained at the W. M. Keck Observatory, which is operated by the
University of California and the California Institute of Technology, and at the National Optical Astronomy Observatory, which is operated by the Association of Universities for Research in Astronomy (AURA) under cooperative agreement with the National Science Foundation.
  
This paper uses observations obtained with facilities of the Las Cumbres Observatory Global Telescope.

{\it Facilities:} \facility{Kepler}.

\clearpage

\begin{deluxetable}{lcccccccccc}
\tabletypesize{\scriptsize}
\tablecaption{Key Properties of Planets and Planet Candidates}
\tablewidth{0pt}
\tablehead{
\colhead{KOI} & 
\colhead{Epoch\tablenotemark{a}} & 
\colhead{$P$} & 
\colhead{$T_{\rm Dur}$} & 
\colhead{$R_p$\tablenotemark{b}} & 
\colhead{$a$\tablenotemark{b}} & 
\colhead{nTT\tablenotemark{c}} & 
\colhead{$\sigma_{TT}$} & 
\colhead{RMS} & 
\colhead{MAD\tablenotemark{d}} & 
\colhead{$M_{\rm p,max}$\tablenotemark{e}} \\ 
\colhead{   } & 
\colhead{(d)} & 
\colhead{(d)} & 
\colhead{(hr)} & 
\colhead{($R_{\oplus}$}) & 
\colhead{(AU)} & 
\colhead{}    & 
\colhead{(d)} & 
\colhead{(d)} &
\colhead{(d)} &
\colhead{($M_{\rm Jup}$)} 
   }
\startdata
168.03=Kepler-23 b &  71.3022 &   7.1073 &  4.77 &   1.9 & 0.075 &           65 & 0.0328 &   0.0565 &   0.0268 &   0.8 \\ 
168.01=Kepler-23 c &  66.2926 &  10.7421 &  6.13 &   3.2 & 0.099 &           44 & 0.0091 &   0.0133 &   0.0090 &   2.7 \\ 
168.02 &  80.5655 &   15.275 &  5.73 &   2.2 & 0.125 &           32 & 0.0443 &   0.0286 &   0.0109 & \nodata \\ 
1102.04 &  70.0712 &   4.2443 &  2.43 &   1.7 & 0.052 &          103 & 0.0417 &   0.0427 &   0.0261 & \nodata \\ 
1102.02=Kepler-24 b &  73.5689 &   8.1453 &  4.02 &   2.4 & 0.080 &           58 & 0.0171 &   0.0307 &   0.0139 &   1.6 \\ 
1102.01=Kepler-24 c &   70.5860 & 12.3335 &  3.71 &   2.8 & 0.106 &           37 & 0.0166 &   0.0239 &   0.0190 &   1.6 \\ 
1102.03 &  77.7512 &  18.9981 &  3.09 &   1.7 & 0.141 &           23 & 0.0184 &   0.0267 &   0.0159 & \nodata \\ 
\enddata
\tablenotetext{a}{BJD-2454900}
\tablenotetext{b}{Updated to reflect stellar properties and dilution from Table \ref{tabStars}}
\tablenotetext{c}{Number of transit times measured in Q0-6}
\tablenotetext{d}{Median absolute deviation from linear ephemeris measured during Q0-6}
\tablenotetext{e}{Based on assumption of dynamical stability and stellar mass from Table \ref{tabStars}}
\label{tabPlanets}
\end{deluxetable}

\begin{deluxetable}{lcccc}
\tabletypesize{\scriptsize}
\tablecaption{Transit Times for Kepler Transiting Planet Candidates during Q0-6}
\tablewidth{0pt}
\tablehead{
\colhead{KOI} & \colhead{n} & \colhead{$t_n$}        & \colhead{TTV$_n$} & \colhead{$\sigma_n$} \\ 
\colhead{}    & \colhead{}      & \colhead{BJD-2454900} & \colhead{(d)}  & \colhead{(d)} 
   }
\startdata
\multicolumn{2}{l}{168.01 = Kepler-23c} & \multicolumn{3}{c}{$        66.292554 + n \times        10.742052$} \\
 168.01 &            0 &   66.2749 & -0.0177 & 0.0089 \\ 
 168.01 &            1 &   77.0227 & -0.0119 & 0.0088 \\ 
 168.01 &            2 &   87.7845 &  0.0078 & 0.0088 \\ 
 168.01 &            4 &  109.2436 & -0.0171 & 0.0092 \\ 
 168.01 &            5 &  119.9877 & -0.0152 & 0.0084 \\ 
\enddata
\tablecomments{
Table \ref{tabTTs} is published in its entirety in the electronic edition of the {\it Astrophysical Journal}.  A portion is shown here for guidance regarding its form and content.
} 
\label{tabTTs}
\end{deluxetable}

\begin{deluxetable}{llccccccccc} 
\tabletypesize{\scriptsize}
\tablecaption{Table of Key Properties of Host Stars}
\tablewidth{0pt}
\tablehead{
\colhead{Kepler} &
\colhead{KOI} & 
\colhead{KIC-ID} & 
\colhead{$Kp$} & 
\colhead{Contam.} &
\colhead{$T_{\rm eff}$} & 
\colhead{[M/H]} & 
\colhead{$\log(g)$} & 
\colhead{$M_{\star}$} & 
\colhead{$R_{\star}$} & 
\colhead{Sources\tablenotemark{a}} \\
\colhead{}    & 
\colhead{}    & 
\colhead{}    & 
\colhead{}    & 
\colhead{}    & 
\colhead{(K)} & 
\colhead{}    & 
\colhead{(cgs)} & 
\colhead{($M_{\odot}$)} & 
\colhead{($R_{\odot}$)}  & 
\colhead{} 
}
\startdata
\KepNumXXX & 168 & 11512246	& 13.438 & 0.025 & $5760(124)$ & $-0.09(14)$ & $4.00(14)$ & $1.11^{+0.09}_{-0.12}$ & $1.52^{+0.24}_{-0.30}$ & M,N,B11 \\
\KepNumYYY & 1102 & 3231341 & 14.925 & 0.086 & 5800(200)   & -0.24(0.40) & 4.34(30) & $1.03^{+0.11}_{-0.14}$ & $1.07^{+0.16}_{-0.23}$ & B11 \\

\enddata
\tablenotetext{a}{Sources for stellar properties.  Spectroscopic stellar parameters are from: 
B11=Borucki \etal (2011), M=McDonald Observatory, N=Nordic Optical Telescope. 
Stellar masses and radii based on comparison to Yonei-Yale models (Demarque \etal 2004). 
Quoted uncertainties do not include systematic uncertainties due to stellar models.  
 }
\label{tabStars}
\end{deluxetable}

\begin{deluxetable}{rrccccccccc} 
\tabletypesize{\scriptsize} 
\tablecaption{Table of Statistics for Pairs of Neighboring Planets Candidates} 
\tablewidth{0pt} 
\tablehead{ 
\colhead{KOI$_{\rm in}$} &  
\colhead{KOI$_{\rm out}$} & 
\colhead{$P_{\rm out}/P_{\rm in}$}  & 
\colhead{$\frac{R_{p,in}}{R_{p,out}}$} & 
\colhead{$C$}  & 
\colhead{FAP$_{\rm TTV,C}$}  & 
\colhead{$\xi$}  & 
\colhead{$\xi_{5/95}$}  & 
\colhead{$\frac{\mathrm{RMS_{in}}}{\mathrm{RMS_{out}}}$}  & 
\colhead{$\frac{\mathrm{MAD_{in}}}{\mathrm{MAD_{out}}}$}  & 
\colhead{$\kappa$}   
%
   } 
\startdata
168.03 & 168.01 &  1.511 &  0.54 & -0.863 & $<10^{-3}$  & 0.89 & 0.57 &   4.27 &   2.35 &   1.02  \\ 
244.02 & 244.01 &  2.039 &  0.58 & -0.774 & $<10^{-3}$  & 1.57 & 2.62 &   1.74 &   2.69 &   0.54  \\ 
250.01 & 250.02 &  1.404 &  1.01 & -0.825 & $<10^{-3}$  & 1.53 & 2.55 &   1.22 &   0.96 &   1.56  \\ 
738.01 & 738.02 &  1.286 &  1.15 & -0.692 & 0.0010     & 0.90 & 0.45 &   0.57 &   0.66 &   1.17  \\ 
806.03 & 806.02 &  2.068 &  0.26 & -0.841 & $<10^{-3}$  & 0.86 & 0.74 &  13.42 &  27.81 &   0.54  \\ 
841.01 & 841.02 &  2.043 &  0.81 & -0.803 & $<10^{-3}$  & 0.88 & 0.42 &   1.02 &   2.34 &   0.59  \\ 
870.01 & 870.02 &  1.520 &  1.05 & -0.223 & 0.2972     & 0.89 & 0.46 &   0.49 &   0.56 &   1.09  \\ 
935.01 & 935.02 &  2.044 &  1.14 & -0.237 & 0.3140     & 0.99 & 0.40 &   0.80 &   1.56 &   1.01  \\ 
952.01 & 952.02 &  1.483 &  0.98 & -0.728 & $<10^{-3}$  & 1.12 & 2.38 &   0.67 &   0.43 &   0.79  \\ 
1102.02 & 1102.01 &  1.514 &  1.14 & -0.905 & $<10^{-3}$  & 1.11 & 2.91 &   1.28 &   0.73 &   1.25  \\ 
\enddata
\label{tabPlanetPairs}
\end{deluxetable}



\end{document}